\documentclass[12pt,twocolumn,a4paper]{aastex63}
\hypersetup{linkcolor=cyan,citecolor=blue,urlcolor=cyan}

\usepackage{xspace}

\newcommand{\cdig}{C$_{DIG}$\xspace}

\newcommand{\wha}{equivalent width\xspace}

\newcommand{\Q}{ionization parameter\xspace}

\newcommand{\hii}{H\,\textsc{II}\xspace}

\newcommand{\Z}{metallicity\xspace}

\newcommand{\doha}{$\rm \Delta log[O\textsc{I}]/H\alpha$\xspace}

\newcommand{\sha}{[S\textsc{ii}]/H$\alpha$\xspace}
\newcommand{\shar}{[S\textsc{ii}]/H$\alpha$ ratio\xspace}

\usepackage{float}
\usepackage[caption = false]{subfig}
\usepackage{xcolor}

\shorttitle{diffuse ionized gas in gas-stripped galaxies}
\shortauthors{Tomi\v{c}i\'{c} et al.}

\graphicspath{{./}{figures/}}

\begin{document}

\title{GASP XXXV: Characteristics of the diffuse ionised gas in  gas-stripped  galaxies}

\correspondingauthor{Tomi\v{c}i\'{c} Neven}
\email{neven.tomicic@inaf.it}

\author{Tomi\v{c}i\'{c} Neven}
\affiliation{INAF- Osservatorio Astronomico di Padova, Vicolo Osservatorio 5, 35122 Padova, Italy}

\author{Vulcani Benedetta}
\affiliation{INAF- Osservatorio Astronomico di Padova, Vicolo Osservatorio 5, 35122 Padova, Italy}

\author{Poggianti  Bianca M.}
\affiliation{INAF- Osservatorio Astronomico di Padova, Vicolo Osservatorio 5, 35122 Padova, Italy}

\author{Werle Ariel}
\affiliation{INAF- Osservatorio Astronomico di Padova, Vicolo Osservatorio 5, 35122 Padova, Italy}

\author{M\"uller Ancla}
\affiliation{Ruhr-Universit\"at Bochum, Faculty of Physics and Astronomy, Astronomical Institute, Universit\"atsstra\ss{}e 150, 44801 Bochum, Germany }

\author{Mingozzi Matilde}
\affiliation{Space Telescope Science Institute, 3700 San Martin Drive, Baltimore, MD 21218, USA }

\author{Gullieuszik  Marco}
\affiliation{INAF- Osservatorio Astronomico di Padova, Vicolo Osservatorio 5, 35122 Padova, Italy}

\author{Wolter Anna}
\affiliation{INAF- Osservatorio Astronomico di Brera, via Brera 28, 
20121 Milano, Italy  }

\author{Radovich Mario}
\affiliation{INAF- Osservatorio Astronomico di Padova, Vicolo Osservatorio 5, 35122 Padova, Italy}

\author{Moretti Alessia}
\affiliation{INAF- Osservatorio Astronomico di Padova, Vicolo Osservatorio 5, 35122 Padova, Italy}

\author{Franchetto Andrea}
\affiliation{Dipartimento di Fisica e Astronomia ``Galileo Galilei”, Universit\`a di Padova, vicolo dell’Osservatorio 3, IT-35122, Padova, Italy}
\affiliation{INAF- Osservatorio Astronomico di Padova, Vicolo Osservatorio 5, 35122 Padova, Italy}

\author{Bellhouse Callum}
\affiliation{INAF- Osservatorio Astronomico di Padova, Vicolo Osservatorio 5, 35122 Padova, Italy}

\author{Fritz Jacopo}
\affiliation{Instituto de Radioastronom\'ia y Astrof\'isica, UNAM, Campus Morelia, A.P. 3-72, C.P. 58089, Mexico}

%\author[0000-xxx]{Name}
%\affiliation{INAF}

\begin{abstract}

The diffuse ionized gas (DIG) is an important component of the interstellar medium that can provide insights into the different physical processes affecting the gas in galaxies. We utilise optical IFU observations of 71 gas-stripped and control galaxies from the Gas Stripping Phenomena in galaxies (GASP) survey, to analyze the gas properties of the dense ionized gas and the DIG, such as metallicity, ionization parameter $\log(q)$, and the difference between the measured $\log[O\textsc{i}]/H\alpha$ and the value predicted by star-forming models, given the measured  $\log[O\textsc{iii}]/H\beta$ ($\rm \Delta log[O\textsc{i}]/H\alpha$). We compare these properties at different spatial scales, among galaxies at different gas-stripping stages, and between disks and tails of the stripped galaxies. The metallicity is similar between the dense gas and DIG at a given galactocentric radius. The $\log(q)$ is lower for DIG compared to dense gas. The median values of $\log(q)$ correlate best with stellar mass, and the most massive galaxies show an increase in $\log(q)$ toward their galactic centers. The DIG clearly shows higher $\rm \Delta log[O\textsc{i}]/H\alpha$ values compared to the dense gas, with much of the spaxels having LIER/LINER like emission. The DIG regions in the tails of highly stripped galaxies show the highest $\rm \Delta log[O\textsc{i}]/H\alpha$, exhibit high values of $\log(q)$ and extend to large projected distances from star-forming areas (up to 10 kpc). We conclude that the DIG in the tails is at least partly ionized by a process other than star-formation, probably by mixing, shocks and accretion of inter-cluster and interstellar medium gas. 

\end{abstract}


%\maketitle

%%%%%%%%%%%%%%%%%%%%%%%%%%%
%%%%
%%   INTRODUCTION
%%%%
%%%%%%%%%%%%%%%%%%%%%%%%%%%

\section{Introduction} \label{sec:intro}

The diffuse ionized gas (DIG) is a crucial but still poorly understood component of the interstellar medium (ISM) of galaxies.
This gas phase is distributed across large spatial scales (up to few kpc compared to $\approx$ 50\,pc scales of star-forming associations; \citealt{Levy19}), with lower gas densities ($\rho \sim 10^{-1}$ cm$^{-3}$) and higher  temperatures ($\gtrsim10^4$ K) than \hii regions\footnote{\hii regions have temperatures of $\approx10^4$ K and electron densities of $\rm \approx 100\, cm^{-3}$ (\citealt{HummerStorey87}, \citealt{Scaife13}).} (\citealt{Collins01}, \citealt{Reynolds01}, \citealt{Haffner09}, \citealt{Barnes14} \citealt{Bruna20}). 
Other defining characteristics of the DIG include lower surface brightness of the Balmer emission line H$\alpha$, higher \sha and [N\textsc{ii}]/H$\alpha$  line ratios ($\rm  [S\textsc{ii}]\,\lambda6717,6731/H\alpha>0.2$) compared to the HII regions (\citealt{Reynolds84}, \citealt{Reynolds92}, \citealt{Martin91},   \citealt{Madsen06}, \citealt{Tomicic17}, \citealt{Kreckel16}, \citealt{Kumari19}, \citealt{Levy19}).

There is no consensus in the literature about what are the main sources of ionization in the DIG. 
The process is likely manifold, and a number of sources have been proposed. 
In highly star-forming galaxies, the DIG can be mostly ionized by photons escaping from \hii regions (\citealt{Reynolds92}, \citealt{Minter98}, \citealt{Haffner09}, \citealt{Relano12}). 
However, the picture is more complex in early-type galaxies or in the bulges of spirals.
In early-type galaxies with detected emission lines, where the DIG usually comprises most of the ISM,  photons from hot, low-mass evolved stars (HOLMES; \citealt{Flores-Fajardo2011}, \citealt{Singh13},  \citealt{Belfiore17b},  \citealt{Zhang17}) can generally account for all of the observed $H\alpha$ emission \citep{Cid11}, with the rare exception of early-type galaxies undergoing rejuvenation events \citep{Herpich18, Werle20}, where \hii regions may contribute to the ionizing field.
Using data from the CALIFA survey (Calar Alto Legacy Integral Field Area), \cite{Lacerda18} found ionization consistent with HOLMES also in spirals, especially at low galactocentric distance.
Other possible sources include supernova shocks and turbulence (\citealt{Slavin93}, \citealt{Minter97},  \citealt{Otte02}, \citealt{Hoopes03}), along with  magnetic reconnection (\citealt{Raymond92}),  collisional excitation from the electrons scattered from dust grains (\citealt{Weingartner01}) and cosmic rays (\citealt{Barnes14}). 
Furthermore, some observations indicate that the emission line ratios and temperature typical of the DIG could be produced by a turbulent mixing of the different hot and cold layers occuring as an aftermath of the interaction of the intra-cluster medium (ICM) and ISM 
(\citealt{CowieSongaila77}, \citealt{Slavin93}, \citealt{Binette09}, \citealt{Fumagalli14}, \citealt{Fossati16}, \citealt{Consolandi17}, \citealt{Campitiello21}, \citealt{Muller21}).

DIG contributes with a fraction of 20\,\% to 90\,\% of the total H$\alpha$ flux in galaxy disks, with a mean fraction around $50\,\%-60\,\%$ (\citealt{Hoopes03}, \citealt{Oey07}, \citealt{Sanders17}, \citealt{Tomicic17}, \citealt{Poetrodjojo19}, \citealt{Bruna20}, \citealt{Tomicic21a}). 
This large contribution may cause star-formation rates (SFRs) to be overestimated, as H$\alpha$ flux from the DIG may be wrongly associated with star-formation. 
There is a debate about to what extent does the DIG affect measurements of the gas-phase metallicity and its radial slope (\citealt{Searle71}, \citealt{Costas92}, \citealt{Sanchez14}, \citealt{Belfiore17},  \citealt{Sanders17}, \citealt{SanchezMeng2018}, \citealt{Zhang17}, \citealt{Asari19}, \citealt{Kumari19}, \citealt{Poetrodjojo19}), as some observations indicate lower metallicity (up to 1 dex) in the DIG compared to  nearby \hii regions.
The DIG may also exhibit different values of line ratios and ionizing parameter $\log(q)$, further affecting observations and analysis of ISM characteristics, as well as adding scatter in the distribution of galaxy properties measured from unresolved observations    (\citealt{Martin91}, \citealt{Flores-Fajardo2011}, \citealt{Dopita14}, \citealt{Zhang17}, \citealt{Poetrodjojo18}, \citealt{Mingozzi20}). 
Furthermore, the detection of gas that shows different line ratios and ionization parameter located at large distances from \hii regions --- larger than the  thickness of a typical galactic disk ($\approx$1 kpc) ---  would indicate that sources other than star-forming (SF) regions are ionizing such gas (for example HOLMES, shocks, or mixing of different gas layers; \citealt{Flores-Fajardo2011}, \citealt{Zhang17}, \citealt{Poetrodjojo18}, \citealt{Poggianti19b}). 
Different galactic characteristics (like mass, SFR, age, etc.) and external physical processes such as galaxy interactions and  gas-stripping caused by ram pressure (\citealt{Toomre72}, \citealt{GunnGott72}), may affect ionization parameter and various line ratios  (\citealt{Nagao06}, \citealt{Maier06}, \citealt{Flores-Fajardo2011},  \citealt{Zhang17}, \citealt{Sanchez20}).

Recent developments in observational astrophysics, especially in the field of Integral Field Unit (IFU) spectroscopy, have helped to probe the physics of the ISM and DIG in galaxies with better spatial and spectral resolutions (\citealt{Slavin93}, \citet{Weingartner01}, \citealt{Hoopes03}, \citet{Binette09}, \citealt{Bundy15}, \citealt{Lacerda18}, \citealt{Sanchez20},  etc.).
In particular, large IFU surveys such as MaNGA (Mapping Nearby Galaxies at Apache Point Observatory, \citealt{Bundy15}) and CALIFA have allowed DIG studies in statistical samples that include a variety of galaxy types.

An IFU survey that stands out due to its sample selection is the GASP project (GAs Stripping Phenomena in galaxies with MUSE; \citealt{Poggianti17}), based on Multi-Unit Spectroscopic Explorer (MUSE, \citealt{Bacon10}) observations.
GASP is a multi-wavelength  survey that studies gas-stripping processes in 114 galaxies spanning the stellar mass range $10^9< M_\ast/M_\odot<10^{11.5}$ at redshift $0.04<z<0.1$ in  clusters (from the  WINGS and OMEGAWINGS surveys; \citealt{Fasano06}, \citealt{Gullieuszik15}) and the field (from the PM2GC catalog; \citealt{Calvi11}). 
The project targets galaxies in a variety of environments (isolated, in filaments, groups and clusters, in junctions of the cosmic web; \citealt{Poggianti17, Vulcani2021}) that are 
subject to different levels of ram-pressure stripping, galactic interaction and evolution.
All of these characteristics make GASP the ideal dataset for studying star-formation and DIG in ram-pressure stripped tails and provide insights on how (and to what extent) the DIG properties are shaped by the physical processes that drive galaxy evolution.

In \cite{Tomicic21a} (hereafter GASP XXXII), we have focused on measuring the fraction of $H\alpha$ flux associated with the DIG (hereafter \cdig) in 71 GASP galaxies.
In this paper, we aim to further analyze the properties of the dense gas and DIG in this galaxy sample (Sec. \ref{sec:Data}). 
First, we will analyze $\log(q)$, defined here as the ratio of the hydrogen-ionizing photon density over the local hydrogen gas density, and \doha defined here as  the horizontal distance of the $\rm O\textsc{i}/H\alpha$ values from the BPT-$\rm [O\textsc{i}]$ line (\citealt{Kewley06}) that separates SF and LINER/LIER regions (Sec. \ref{sec:Data}).
We will use  both  spaxel-by-spaxel (Sec. \ref{subsec:Results, spatial}) and integrated data (Sec. \ref{subsec:Results, GASP integrated}, and \ref{subsec:Results, GASP integrated functions}) of the galaxies. 
Second, following the results by \citet{Zhang17}, we aim to test if the gas-phase metallicity differs between the dense gas and the DIG at a specific galactocentric radius (Sec. \ref{subsec:Results, R variation}). 
Third, we will compare the properties of the dense gas and the DIG in the disk and the tails in order to understand whether ram-pressure stripping affects their properties.
Furthermore, we will check if the distance from SF regions affects variation in the line ratios, as a result of DIG begin ionised by other sources than SF regions (Sec. \ref{subsec:Results, spatial, distance}).
The discussion of the results and our conclusions are written in Sec. \ref{sec: Discussion} and  \ref{sec: Conclusion}, respectively. 

In this paper we adopted standard cosmological constants of H$_0=70$ km$\rm \,s^{-1}Mpc^{-1}$,  $\rm \Omega_M=0.3$, $\Omega_\Lambda=0.7$, and the initial mass function (IMF) from \citet{Chabrier03}.

%%%%%%%%%%%%%%%%%%%%%%%%%%%
%%%%
%%   DATA
%%%%
%%%%%%%%%%%%%%%%%%%%%%%%%%%

\section{Data} \label{sec:Data}

%%%%
%%   Subsec

\subsection{Galaxy sample }\label{subsec:Data, gasp, obs}

For this paper, as in \citealt{Vulcani18b}, we consider galaxies with  gas-stripping features (hereafter referred to as stripping galaxies) and compare them to  undisturbed galaxies that do not show any sign of interactions or gas-stripping on their H$\alpha$ maps (referred to as control sample galaxies).
Galaxies that are tidally interacting with others, or merging, or lopsided, or cosmic web enhanced, etc. are not included in this paper (\citealt{Vulcani17}, \citealt{Vulcani18}, \citealt{Vulcani18a}, \citealt{Vulcani19a}, \citealt{Vulcani20b}, \citealt{Vulcani2021}).

For each galaxy, Poggianti et al. \textit{in prep} have designated a specific number indicating a stage of stripping, referred here as J stage number (Tab. \ref{tab:Tab01_Jstage}). 
The J stage numbers classify non-stripped galaxies with 0, galaxies in early stage of stripping with 0.5, clear gas-stripped galaxies and jellyfishes with 1 and 2, respectively,  galaxies with a truncated $\rm H\alpha$ disk are assigned a J stage number of 3.
The truncated galaxies are galaxies where the gas disk was mostly stripped, except in the galactic center (\citealt{Koopmann04}, \citealt{Fritz17}).
We will use this classification in this paper to compare galaxies at different stripping phases.

\begin{table}
\centering
\caption{ The explanation of  J stage numbers used in this paper (defined by B.M.  Poggianti et al. \textit{in prep}) and number of galaxies with J stage number.   }
\begin{tabular}{ccc}
\\
J stage  & Explanation & Number of galaxies \\ 
\hline 
0 & non-stripped, control galaxies & 30 \\ 
0.5 & early stage of stripping & 10 \\ 
1 & clear gas-stripped galaxies & 11 \\
2 &  jellyfish galaxies & 16 \\
3 & truncated galaxies & 4 \\
\hline 
\end{tabular}
\label{tab:Tab01_Jstage}
\end{table}    

Both, stripping and control samples are drawn from \citet{Vulcani18b}, with some modifications. 
The stripping sample is composed of 41 galaxies. 
With respect to \citet{Vulcani18b}, we exclude JO149 and JO95, since we are not able to measure their effective galactocentric radii and orientation (\citealt{Franchetto20}). 
Furthermore, we add JO93 (a member of the control sample in \citealt{Vulcani18b}) because its H$\alpha$ map indicates an initial phase of stripping. 

The control sample includes 30 galaxies, 16 of which are cluster members and 14 are field galaxies. 
With respect to \citet{Vulcani18b}, we exclude P19482, as \cite{Vulcani19} showed that the galaxy is not undisturbed but undergoing cosmic web enhancement. 
The control sample galaxies by definition have no tails.

%%%%
%%   Subsec

\subsection{Observations, line maps, and galaxy orientation}\label{subsec:Data, line maps}

The galaxies were observed by MUSE, an IFU at ESO-VLT (the Very Large Telescope of the European Southern Observatory), covering the optical range of $4800-9300$\,\AA. 
The data analysis of the observed cubes follows established procedures described in \cite{Poggianti17}. 
Briefly, the observed spectrum in each spaxel of the cube was corrected for the effect of foreground dust extinction of the Milky Way, with corresponding $E_{B-V}$ values for each galaxy in the line of sight (LOS) on sky measured by \cite{SFD}, considering the recalibration introduced by \cite{Schlafly11}, and assuming the extinction curve of \citet{Cardelli89} with $\rm R_V = 3.1$.
To account for seeing effects, the data were smoothed and convolved in the spatial dimension using a $5\times5$ pixel kernel, which corresponds to 1 arcsec, or  $0.7-1.3$\,kpc, depending on the galaxy's redshift. 

The convolved cubes were analyzed with the spectrophotometric code SINOPSIS (\citealt{Fritz17}) to fit the stellar component of the spectra, and then with KUBEVIZ (\citealt{Fossati16}) to fit the gas emission lines in the stellar-continuum subtracted spectra. 
The emission lines used in this work are: $\rm H\beta$, $\rm [O\,\textsc{iii}]\lambda5007$, $\rm [O\,\textsc{i}]\lambda6300$  (referred hereafter as $\rm [O\textsc{i}]$), $\rm H\alpha$, $\rm [N\,\textsc{ii}]\lambda6584$, and $\rm [S\textsc{ii}]\lambda6713,6731$ (referred hereafter as $\rm [S\textsc{ii}]$).

Maps of $\rm H\alpha$ surface brightness corrected for attenuation, labeled as $\rm \Sigma H\alpha,corr$, were calculated assuming the intrinsic Balmer line ratio $\rm H\alpha/H\beta=2.86$, ionized gas temperature of $\rm T\approx10^4$\,K  and case B recombination (\citealt{HummerStorey87}, \citealt{Osterbrock92}).
We applied a cut in signal-to-noise $\rm S/N\geq$4 for all emission lines in our analysis and for our results.

The boundaries of stellar disks were estimated using an isophote in the continuum maps that is 1\,$\sigma$ above the average sky background noise, and the center of the galaxies is designated to be the centroid of the brightest central region in the continuum map (\citealt{Gullieuszik20}).
We define the spaxels (with the ionized gas emission) outside the stellar disks as part of galactic tails. 
The disk inclination with respect to the line of sight and the disk diameter were estimated by \citet{Franchetto20} from the I-band images obtained by convolving the MUSE datacube spectra.

\begin{figure}[t]
\includegraphics[width=0.45\textwidth]{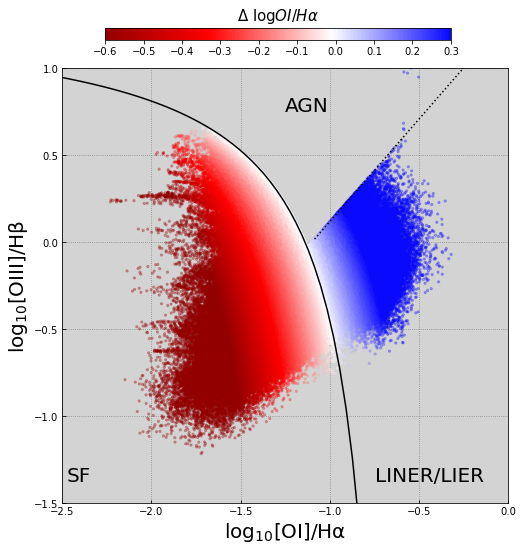}
   \caption{ The BPT-$\rm [O\textsc{i}]$ diagram of the spaxels of all galaxies, color-coded by \doha  values. The \doha  value is defined here as the horizontal distance of the $\rm O\textsc{i}/H\alpha$  values from the BPT-$\rm [O\textsc{i}]$ line (thick black line; \citealt{Kewley06}) that separates SF and LINER/LIER regions. The dashed black line separates AGN regions from LINER/LIER regions (\citealt{Kewley06}).   }
\label{fig:Fig_BPT}
\end{figure}

\begin{figure*}
\centering
 \includegraphics[width=0.95\textwidth]{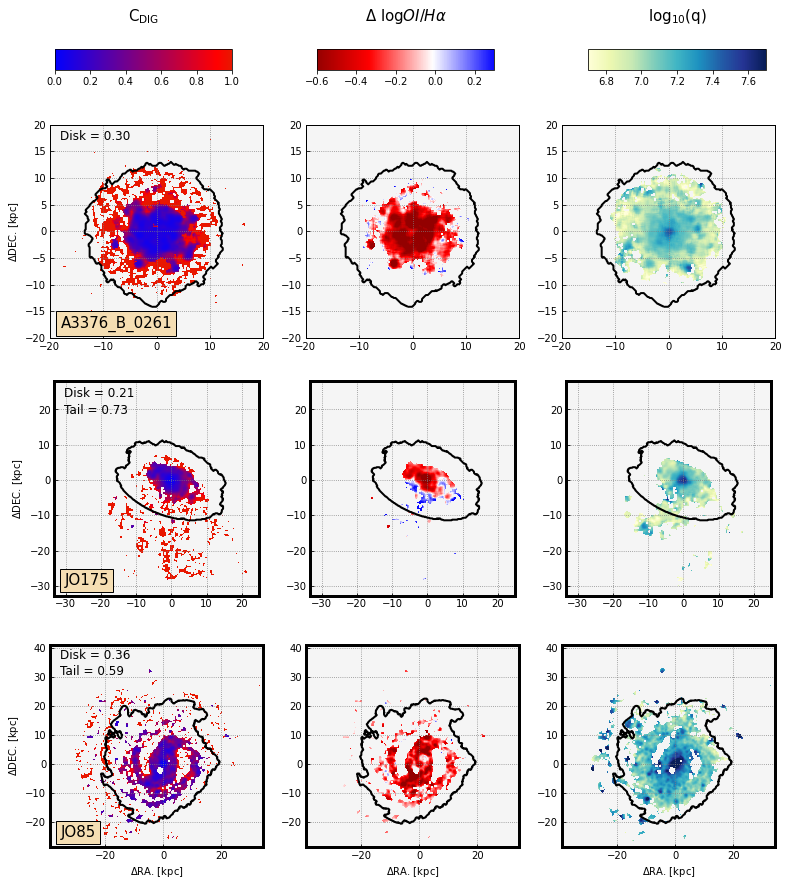}
   \caption{ Galaxies presented here are $A3376\_B\_0261$, JO175, and JO85. The panels showing control sample and stripped galaxies have thin and thick edges, respectively. The ticks on the axes are centered in relation with the galactic center.  We indicate  the galactic disk with the thick, black contour (see text for the definition).  \textit{Left column-} Maps of the diffuse ionized gas fraction ($\rm C_{DIG}$), from \citet{Tomicic21a}. Dense gas  dominated spaxels are blue, while   DIG dominated spaxels are red.  The integrated $\rm C_{DIG}$ values for the disks and tails are labeled in the upper left corner.   \textit{Central column-} Maps of \doha, which  measures distances of $\rm O\textsc{i}/H\alpha$  values from the BPT-$\rm [O\textsc{i}]$ line that separates SF and LINER/LIER regions.  \textit{Right column-} Ionization parameter $\log(q)$  maps. }
    \label{fig:Fig_Cdig_delOIHa_Q_p1}
\end{figure*}

It is important to note that a clear separation between DIG and \hii regions requires a resolution that is not achieved by our observations. 
Thus we can only probe regions where \hii complexes and DIG are mixed together in the line of sight, and measure DIG emission fraction instead.
Therefore, in this paper we use the terms ``dense gas'' and DIG, not \hii regions and DIG.

%%%%
%%   Subsec

\subsection{Metallicity, ionization parameter and \doha}\label{subsec:Data, PYQZ, OIHa}

Two of the main quantities of interest in our analysis are the gas-phase metallicity and ionization parameter. 
These quantities were derived from the emission line ratios, $\rm [O\,\textsc{iii}]\lambda5007$/$\rm [S\textsc{ii}]$ vs. $\rm [N\,\textsc{ii}]\lambda6583$/$\rm [S\textsc{ii}]$,  using the PYQZ code (v0.8.2 version, \citealt{Dopita13}, \citealt{Vogt15}). 
We assume solar metallicity to be $\rm 12+log(O/H)=8.69$.
Details on how we estimate the gas-phase metallicity and $\log(q)$ for the GASP sample are described by \citet{Franchetto20}. 
$\log(q)$ is defined as the logarithm of the ratio of the hydrogen-ionizing photon density (in $\rm cm^{-3}$)  over the local hydrogen gas density (in $\rm cm^{-3}$) multiplied by the speed of light (\citealt{Kewley06}, \citealt{Nagao06}, \citealt{Kewley19}). 
Since it measures the ratio of ionization intensity    and gas density, it may be used as an indication for the ratio of radiation and gas pressures (\citealt{Yeh12}).
We note that PYQZ uses photo-ionization models of line ratios that assume ionization with photons from SF regions.

The Baldwin, Phillips \& Terlevich diagnostic diagrams (BPT; \citealt{Baldwin81}, \citealt{Kewley06}) of emission line ratios are used to infer the dominating ionization mechanism, such as SF, AGN, or LINER\footnote{Low ionization Nuclear Emission Regions (\citealt{Heckman80}).}/LIER\footnote{Low ionization Emission Regions (\citealt{Belfiore16}).}. 
In this work, we only use the BPT diagram based on $\rm [O\,\textsc{i}]\lambda6300$ line (hereafter BPT-$\rm [O\textsc{i}]$, shown in Fig. \ref{fig:Fig_BPT}), and we removed spaxels dominated by AGN ionization.
The $\rm [O\textsc{I}]/H\alpha$ line ratio is sensitive to the hardness of the ionizing radiation field (\citealt{Kewley06}), and LINER/LIER-like spaxels may  potentially indicate that the origin of ionization across the galaxy disks and tails can be other than SF regions. 
 
To quantify the difference in ionization source and variation in gas emission among the galaxies, for each spaxel we measure the difference between the measured $\log[OI]/H\alpha$ and the corresponding value on the  BPT-$\rm [O\textsc{i}]$ line that separates the SF and LINER/LIER region,\footnote{This line was computed as an upper limit of the theoretical
pure stellar photoionization models in the star-burst case (\citealt{Kewley01}, \citealt{Kewley06}).} at the measured $\log$[O\textsc{iii}]/H$\beta$. 
This value is called \doha and is shown in Fig.   \ref{fig:Fig_BPT}. 
Negative \doha values represent regions defined as SF, while positive values represent LINER/LIER like regions.  
We note that negative \doha does not necessarily mean that the ionization is coming only from SF regions, but that these account for the largest fraction.
Similarly, LINER/LIER like regions do not exclude a possibility of partial ionization by star-formation.
The individual BPT diagrams of spatially resolved data for 16 GASP galaxies were shown by \citet{Poggianti19}.

\subsection{Fraction of the diffuse ionized gas}\label{subsec:cdig}

In  GASP XXXII, we estimated the fraction of H$\alpha$ emission coming from DIG for all 71 galaxies studied in this work. 
The initial assumption was that the DIG exhibits lower $\rm H\alpha_{corr}$ flux surface density ($\rm \Sigma H\alpha$) and higher \sha line ratio than the dense gas \citep{Blanc09,Kaplan16}. 
We account for the radial gas-phase metallicity distribution across the galaxy, where we divided the $\rm [S\textsc{ii}]/H\alpha$ ratio by the gas metallicity value at each given radial bin.  
Then we fitted the spaxel by spaxel anti-correlation between $\rm [S\textsc{ii}]/H\alpha$ ratio and the extinction corrected $\rm \Sigma H\alpha_{corr}$, for each individual galaxy.
For this fitting, we designated the spaxels with the highest $\rm [S\textsc{ii}]/H\alpha$ ratio and lowest $\rm \Sigma H\alpha_{corr}$ to have $\rm C_{DIG}=1$, while the data with the highest $\rm \Sigma H\alpha_{corr}$ were assumed to have low $\rm C_{DIG}$ fraction. 
We then derived the relation between the DIG fraction and $\rm \Sigma H\alpha_{corr}$, and created maps of \cdig from the $\rm \Sigma H\alpha_{corr}$ maps.
Further details of the method can be found in GASP XXXII.

In principle, considering metallicity as a parameter in our method to estimate \cdig could introduce a bias in our analysis of \cdig vs. metallicity.
However, at a given radius, metallicity correction  affects \sha of dense gas and DIG spaxels equally, thus not changing the slope of the spaxel by spaxel anti-correlation in the diagrams.
Furthermore, our method uses both \sha and H$\alpha$, thus mitigating strong effects of metallicity on \cdig and lowering the scatter of the data.
Thus, the \cdig values in spaxels at similar radii are not significantly affected by variations in metallicity. 

In what follows, to clearly contrast regions whose emission is generated by different mechanisms, we will consider dense gas dominated spaxels those with $\rm C_{DIG}\leq0.3$, DIG dominated spaxels those with $\rm C_{DIG}>0.7$, and disregard the spaxels with intermediate values.

%%%%%%%%%%%%%%%%%%%%%%%%%%%
%%%%
%%   RESULTS
%%%%
%%%%%%%%%%%%%%%%%%%%%%%%%%%

\section{Results} \label{sec:Results}

In this section we will investigate galaxy metallicity, $\log(q)$ and \doha across our galaxy sample.
We will first consider spatially resolved (i.e. spaxel-by-spaxel) scales, and then the values at larger spatial scales (radial trends and integrated values). 
Furthermore, we will compare i) regions dominated by dense gas emission with those dominated by DIG emission, ii) disks and tails of galaxies, iii) different stages of stripping (J stage) in the stripping sample. 
Lastly, we will investigate trends of $\log(q)$ and \doha with integrated values such as stellar mass, SFR, and metallicity measured at the effective radius.
 Only spaxels ionized by SF processes (according to BPT-[O\textsc{i}]) are considered  when the metallicity and log(q) are analyzed and displayed (spaxel-by-spaxel or integrated values).

%%%%
%%   Subsec

\subsection{Maps of galaxies }\label{subsec:Results, Maps}

We present the maps of $C_{DIG}$, \doha  and $\log(q)$ values for control ($\rm A3376\_B\_0261$) and stripped  galaxies (JO175, JO85) in Fig. \ref{fig:Fig_Cdig_delOIHa_Q_p1}, and  all other galaxies in Appendix \ref{sec:Appendix 1}. 
We indicate the control and stripped sample with thin and thick line edges of panels, respectively. 
We will further quantify the trends of  \doha  and $\log(q)$ as a function of galactocentric radius in Sec. \ref{subsec:Results, R variation}.

The maps of $C_{DIG}$ (left panels of Fig. \ref{fig:Fig_Cdig_delOIHa_Q_p1} and the figures Appendix) reveal some interesting trends: in most cases, the dense gas is confined to the galaxy's core and spiral arms, and the DIG emission is preferentially found between the arms,  in the galaxy's outskirts, and tails of stripped galaxies. 
The maps are lacking \doha values (middle panels) in many spaxels compared to $C_{DIG}$ map due to a low $\rm S/N$ of $\rm [O\textsc{i}]$ emission line.  

Comparing the maps of $C_{DIG}$ and those of \doha and $\log(q)$, we recover a tendency for the dense gas emission dominated areas to have higher ionization parameter and lower \doha than the DIG dominated areas.
Moving from galaxy cores to the outskirts and tails, the incidence of LINER/LIER like spaxels (higher \doha values) increases and $\log(q)$ decreases.

\begin{figure*}[t]
\centering
 \includegraphics[width=1.\textwidth]{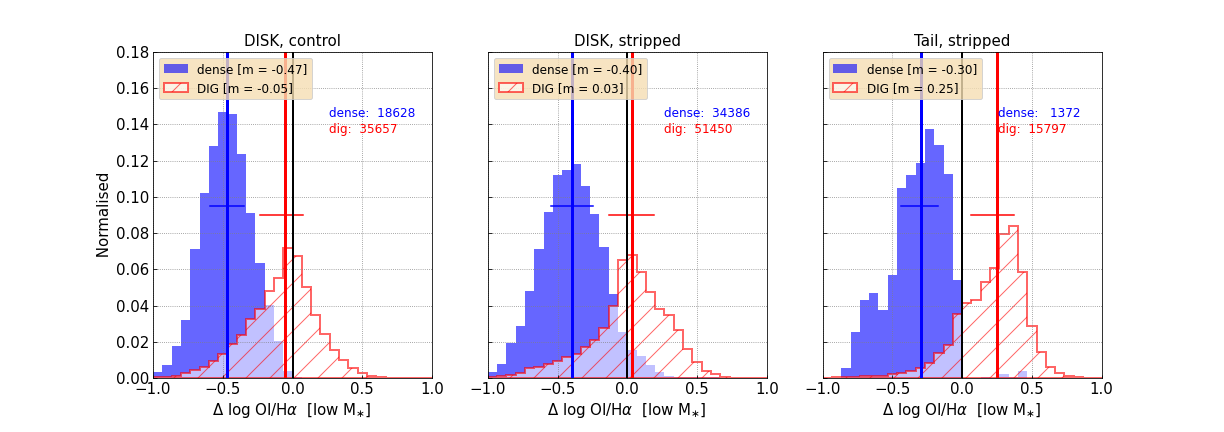}
   \caption{  Histograms of spaxel-by-spaxel  comparison of  \doha values for dense gas (filled, blue) and DIG (hatched, red) across the galaxies. There are 244479 spaxels in total. The data are separated according to location from where they were sampled: disks of the control sample (left), disks (middle) and tails (right) of the stripped galaxies.  The corresponding  median values and the range between first and third percentile  are indicated by vertical and horizontal blue (dense gas) and red (DIG) lines, and they are labeled in the legend of the panels. The numbers of spaxels in corresponding distributions are labeled in the upper right corner of the panels.    }
    \label{fig:Fig_Histo_spatial}
\end{figure*}

%%%%
%%   Subsec

\subsection{Spaxel-by-spaxel comparison }\label{subsec:Results, spatial}

Figure  \ref{fig:Fig_Histo_spatial} compares the  spatially resolved \doha for dense gas dominated spaxels (blue, filled histograms) and DIG dominated spaxels (red, hatched histogram), for different regions (disks and tails) of the control and stripped galaxies.
We remind readers that control sample galaxies have no tails by definition. 
The median values of the corresponding distributions are represented by the blue and red vertical lines for the dense gas and DIG dominated spaxels, respectively.

 While the dense gas in  disks of the control galaxies shows median \doha$\approx -0.5$,  disks and tails of the stripped galaxies have median \doha of  $\approx -0.4$ and $\approx -0.3$ respectively.
In general,  the DIG  dominated spaxels have $\approx 0.4$\,dex higher median \doha  than dense gas areas. 
The DIG spaxels in the disks show more LINER like spaxels (approximately half of spaxels), while the stripped tails show majority (83\%) of spaxels with positive  \doha.

\begin{figure*}[t]
\centering
 \includegraphics[width=1.\textwidth]{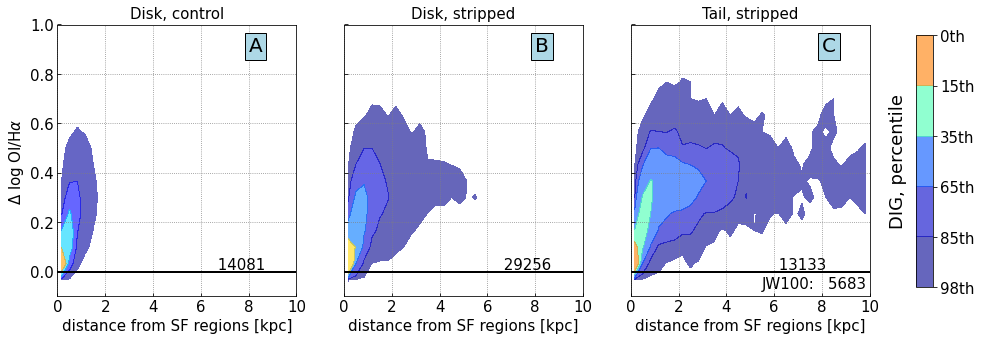}
   \caption{Spaxel-by-spaxel comparison of \doha (y-axis) and the distance from the SF regions  (x-axis) for DIG dominated spaxels in the galaxies.  The contours represent the 15th, 35th, 65th, 85th and 98th percentiles of their spaxel distributions. The numbers of spaxels in corresponding distributions are labeled in the bottom right corner of the panels. We separately included the number for JW100.}
    \label{fig:Fig_Histo_spatial_dist}
\end{figure*}

%%%%
%%   Subsec

\subsection{\doha and  the distance from SF regions }\label{subsec:Results, spatial, distance}

In star-forming galaxies, the DIG can reach scale-heights of $1-2$\,kpc (\citealt{Reynolds84}, \citealt{Haffner09}, \citealt{Bocchio16}, \citealt{Tomicic17}), larger than the thickness of their stellar disks (up to a few hundreds of pc), favoring the hypothesis that it could require ionization by sources other than star-formation (i.e. older stellar populations, or mixing of ISM and ICM). 
Detecting the DIG at large distances from SF regions would validate that hypothesis. 

In this section we consider only the DIG dominated spaxels and characterize their projected distance\footnote{Due to the uncertainty of projections of lengths, this measured distance is a lower limit.} from SF regions (designated by BPT-$\rm [O\textsc{i}]$ diagram). 
For each individual non-SF spaxel, we measure the mean distance from the closest 10 SF spaxels.  
Fig. \ref{fig:Fig_Histo_spatial_dist} presents  \doha values of spaxels in all galaxies as a function of the projected distance from the SF spaxels. 

As in Fig. \ref{fig:Fig_Histo_spatial}, we show data for different types of galaxies (control vs. stripped), their different parts (disks vs. tails).
In general, we notice a positive trend between \doha and distance.
In the stripped tails and in some parts of the stripped galaxies disks,  there are regions that span large distances from the SF spaxels ($>$2\,kpc, and up to 10\,kpc), and which have a mean \doha value of  $\approx 0.4$. 
These spaxels, which exhibit high \doha $\approx 0.4$ and are distant from the SF regions, correspond to highly stripped galaxies (JO93, JW100, JO204, JO194, JO147, JO85, JO60, JO27), of which the edge-on JW100 with its long tail contributes the most (approximately half of those spaxels). 

From these results, we conclude that the DIG in stripped galaxies extends to very large distances from SF regions, which would make its ionization by SF regions alone less probable.

\begin{figure*}
\centering
 \includegraphics[width=1.\textwidth]{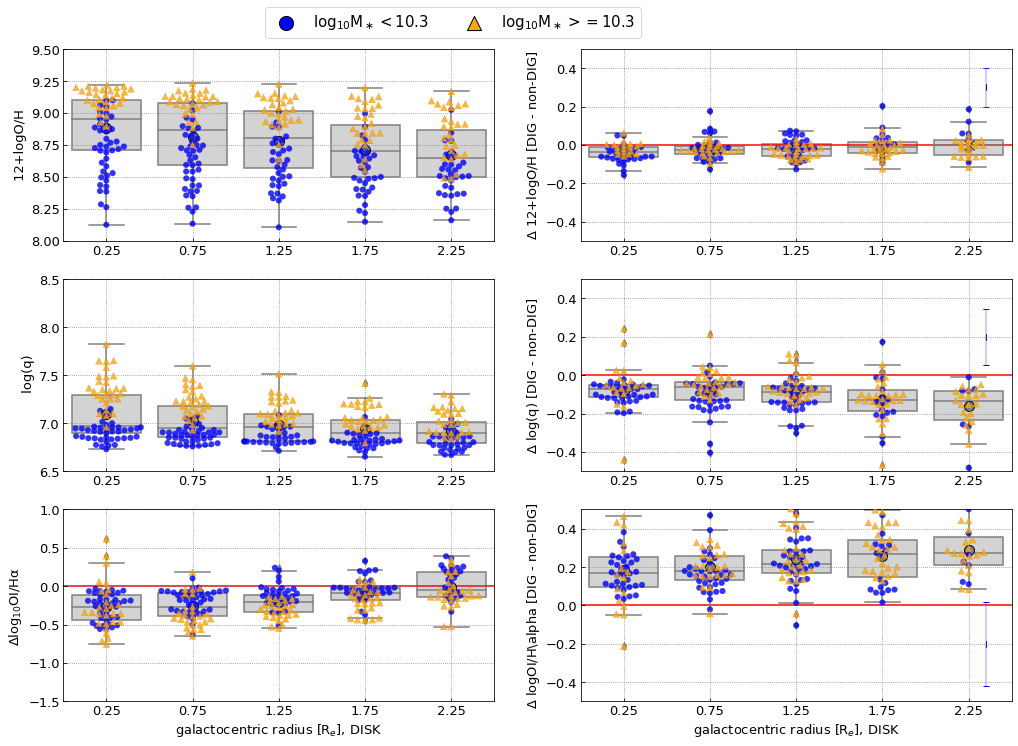}
   \caption{Median values of quantities of spaxels in the disk of each galaxy, shown as absolute values in left panels, and as difference between DIG ($\rm C_{dig}>0.7$) and dense gas  ($\rm C_{dig}<0.3$) dominated spaxels in right panels. Quantities presented here are: gas-phase metallicity (upper panels), $\log(q)$ (middle panels) and \doha values (bottom panels). The values are plotted  as a function of the galactocentric radius (normalised by the effective radius $\rm R_e$).  Each galaxy is represented by a data point, while the quartiles and extent of all data are plotted with grey boxes and vertical lines. We color-code galaxies by their stellar disk mass, $\rm log M_\ast <10.3$ in blue circles and $\rm log M_\ast \geq 10.3$ in orange triangles.    The mean uncertainty  is plotted as  the blue error bar on the right.    }
    \label{fig:Fig_diffZ}
\end{figure*}

%%%%
%%   Subsec

\subsection{Comparing DIG and dense gas properties at different galactocentric radii }\label{subsec:Results, R variation}

Next, we test if regions dominated by the DIG and dense gas emission in disks of both control and stripped galaxies are also characterized by different values of gas-phase metallicity, $\log(q)$, and \doha, at each given galactocentric distance.

We divided the spaxels into annuli of different deprojected galactocentric radii (normalized by the effective radius of the disk;  \citealt{Franchetto20}). 
In each radial annulus, we  measured the median of all  quantities, as well as differences between median values of the DIG and dense gas dominated spaxels. 
We present in  Fig.  \ref{fig:Fig_diffZ} the absolute values (left panels) and the differences  between  median values of dense gas and DIG (right panels) as a function of radial annuli.
The figure shows each galaxy separately as a data point, color-coded by its stellar mass (blue for $\rm log M_\ast <10.3$, and orange for $\rm log M_\ast \geq10.3$).

For the gas-phase metallicity (upper panels), the absolute values decrease with radius, as expected from the fact that metallicity gradients are negative for the great majority of galaxies (\citealt{Mingozzi20}, A. Franchetto et al. submitted). 
The low-mass galaxies have lower metallicities, following the known mass-metallicity relation \citep{Franchetto20}.
A small  difference ($<0.05$ dex) between the DIG and dense gas dominated spaxels is detected, but no variation of this difference with galactocentric radius is found. 
The scatter of the data ($\approx 0.5$ dex) and the uncertainty of individual data points ($\approx 0.1$ dex) are larger than difference in metallicities between dense gas and DIG.

The ionization parameter (middle panels) does not show a clear trend with radius for the low-mass galaxies, but shows a clear decrease with radius for the high-mass galaxies. 
The difference in $\log(q)$ between DIG and dense gas areas shows a negative  offset, with DIG data having $\approx0.1$\,dex lower values. 
Furthermore, the $\log(q)$ offset between DIG and dense gas increases with galactocentric radius. 

Mean absolute values of \doha (bottom panels) increase with radius for galaxies of all masses by $\approx 0.2$\,dex. 
At a given radius, the high-mass galaxies tend to have slightly lower values compared to the low-mass galaxies. 
The offset in \doha indicates that the DIG dominated regions have $0.2-0.3$\,dex higher \doha values compared to the dense gas dominated regions, and the mean value of this difference increases with radius.  

These results indicate that metallicity is not affected by the DIG fraction. The median $\log(q)$ of total gas increases in the centers of high-mass galaxies, with DIG fraction playing  a small role on $\log(q)$. 
The median \doha of total gas is increasing toward the edges of all galaxies, with DIG fraction potentially playing an important role.

\begin{figure*}
\centering
\includegraphics[width=1.\textwidth]{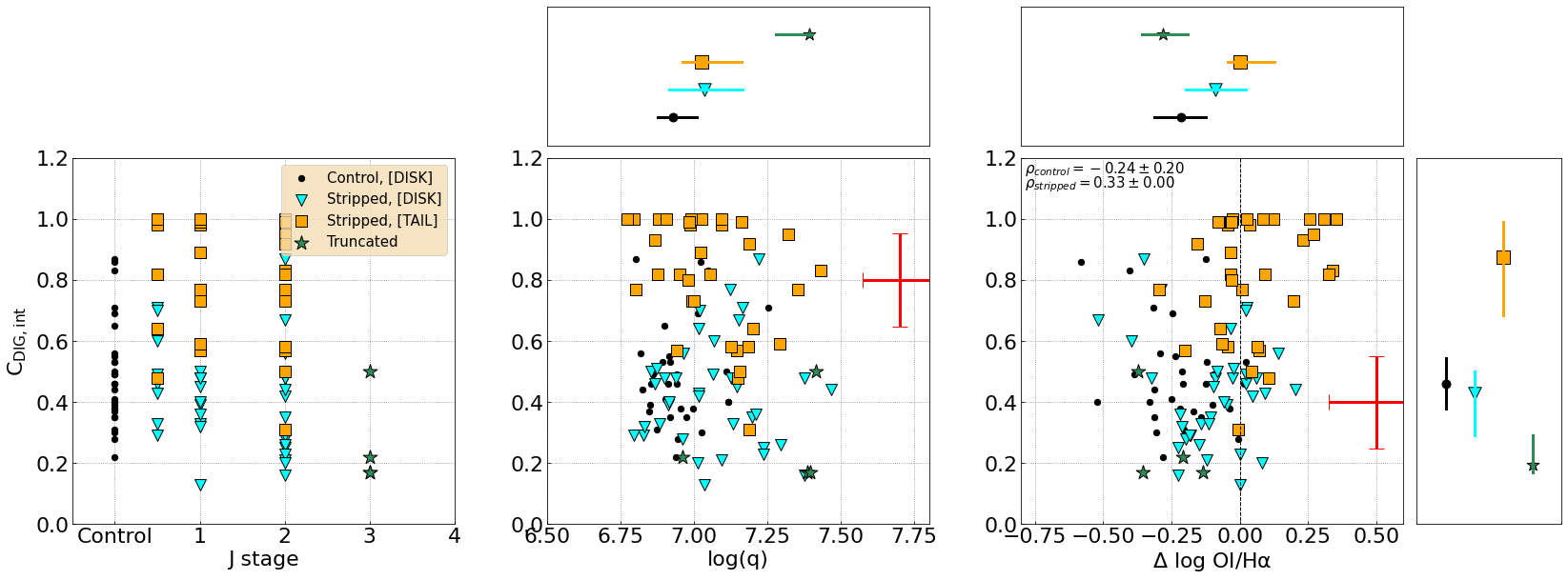}
\caption{  Diagrams of the integrated fraction of the DIG emission ($\rm C_{DIG,int}$) as a function of J stage (left panel), median galactic values of $\rm log_{10}q$ (middle panel) and [O\textsc{i}] excess (right panel). We show control sample  disks (black dots), disks (cyan triangles) and tails (yellow squares) of the stripped galaxies, and truncated disks  (green stars). The mean uncertainty of all values are shown in the right corners of panels, and Pearson coefficients in the upper left corner of right panel.  Above and on the right, we show median values and the first and third quartile of the  data samples in diagrams with their corresponding symbols.}
    \label{fig:Fig_Galaxies int}
\end{figure*}

%%%%
%%   Subsec

\subsection{Galaxy integrated values}\label{subsec:Results, GASP integrated}

We investigate in Fig. \ref{fig:Fig_Galaxies int}  how different physical properties such as the integrated fraction of the DIG emission\footnote{The integrated fraction here is defined as the sum  of the $\rm H\alpha$ luminosity from DIG dominated spaxels over the total $\rm H\alpha$ luminosity of galaxy.} ($\rm C_{DIG,int}$, GASP XXXII) as a function of J stage (left panel), median $\log(q)$ (middle panel) and \doha (right panel).  
We estimated the $\rm C_{DIG,int}$ within the disks and tails of all galaxies (control sample and stripped galaxies).
In the case of median $\log(q)$, we use the median value of only the spaxels designated as SF on the BPT-[O\textsc{i}] diagram.

As seen in GASP XXXII, the disks of stripped and control sample galaxies cover a similar range in $\rm C_{DIG,int}$ (from 0.1 to 0.9), with a median value\footnote{We show median values and first and third quartiles in diagrams above and on the right from the panels.} $\rm C_{DIG,int}=0.44^{|_{0.33}^{0.52}}$  for the entire sample. 
The tails of stripped galaxies have higher DIG fractions (median $\rm C_{DIG,int}=0.83^{|_{0.69}^{0.99}}$) than the disks. 

There is generally no systematic variation of the $\rm C_{DIG,int}$ of the disks with J stage. 
The truncated disks (J=3) show a low DIG fraction ($\approx0.2$), probably due to a large fraction of gas being stripped away from the disk, and preservation of only the dense gas, star-forming regions in their centers.  

Focusing on the ionization parameter, we find that control galaxies have lower median values ($\rm  \log(q)=6.93^{|_{6.88}^{7.0}}$) compared to the stripped galaxies (median $\rm  \log(q)=7.03^{|_{6.94}^{7.17}}$). 
The disks and tails of the stripped sample also show the same median values.
We note that the control sample spans a narrower range of  $\log(q)$  than the disks and tails of stripped galaxies.

The gas in disks of control and stripped sample galaxies has mostly negative \doha values (median $\rm -0.22^{|_{-0.31}^{-0.12}}$ and $\rm -0.09^{|_{-0.2}^{0.02}}$, respectively).
The tails have higher values in \doha than the disks with more than half of them having positive values (median $\rm \Delta\,log_{10}[O\textsc{i}]/H\alpha =0.0^{|_{-0.06}^{0.12}}$).

We also notice a weak correlation between \doha and $\rm C_{DIG,int}$ across  stripped galaxies (Pearson coefficient $\rho=0.33\pm0.00$), while the control galaxies show a slight anti-correlation ($\rho=-0.24\pm0.2$). 
This effect is mostly due to the tails of high $\rm C_{DIG,int}$, which have slightly higher \doha.

\begin{figure*}
\centering
\includegraphics[width=1.\textwidth]{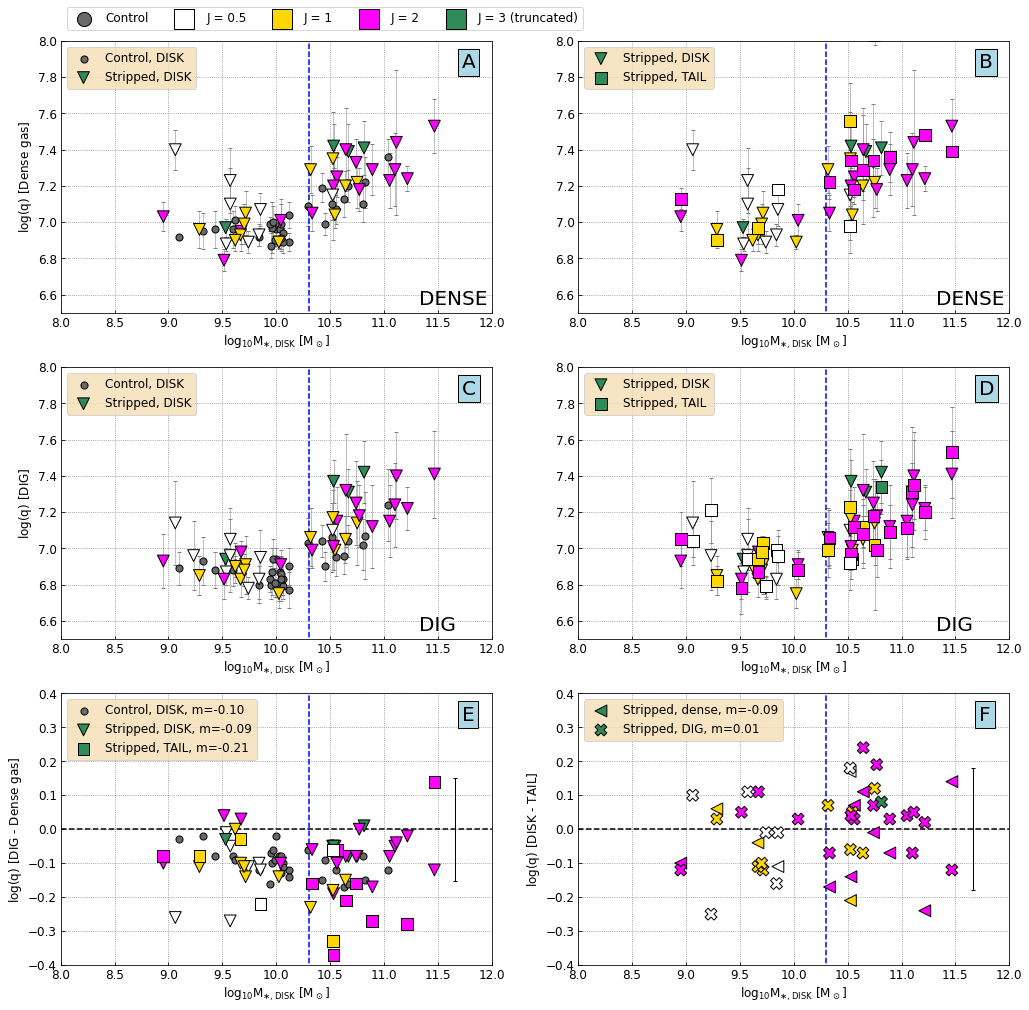}
\caption{Panels A, B, C, and D show integrated (median of spaxels) values of $\log(q)$  as a function of galaxy stellar  mass. Data points are color-coded by J stage (legend on the top).  On the left panels (A and C) we compare disks of the control and stripped samples (with added tails in panel E), while on the right (panels B, D, and F) we compare the disks and tails of only stripped galaxies. We also separately present medians of the dense gas dominated regions (panels A and B)  and medians from the DIG dominated areas (panels C and D).   In panel E, we show the difference between the DIG and dense gas values, while in panel F we show the difference between the disks and tails of the stripped galaxies. The mean of shown values in panels E and F are labeled in the legends for different parts of galaxies.   The separation between low and high mass galaxies  at $\rm log_{10}M_{\ast}=10.3$, used in previous sections, is indicated by the blue, dashed line.  The scatter of spaxel values for each galaxy is presented with error-bars, while mean of data's error-bars in panels E and F is presented with a single error-bar on the right.    }
    \label{fig:Fig_stats_integrated}
\end{figure*}

\begin{figure*}
\centering
\includegraphics[width=1.\textwidth]{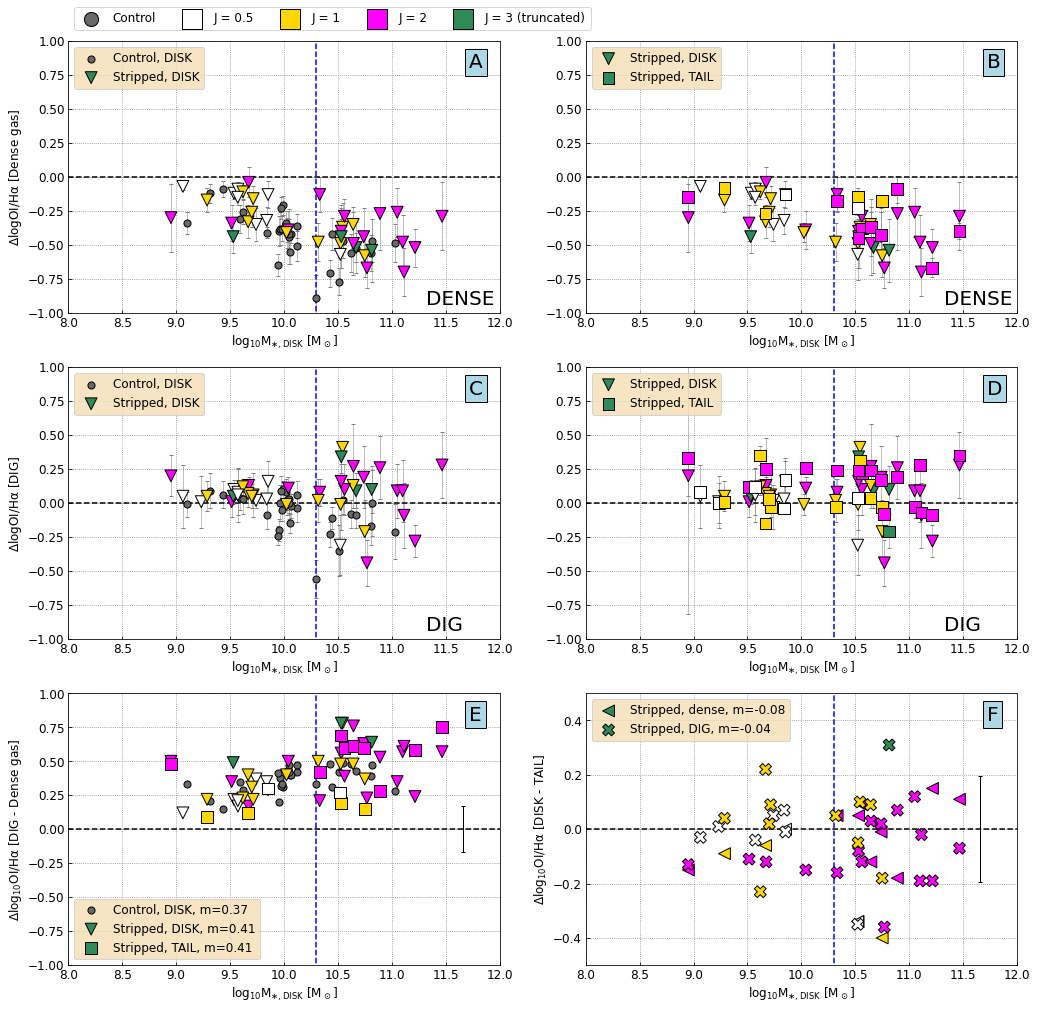}
\caption{Same as in Fig. \ref{fig:Fig_stats_integrated}, but for \doha values, instead of $\log(q)$.     }
    \label{fig:Fig_stats_integrated_p2}
\end{figure*}

These results show that the gas-stripping process, in general, slightly increases the $\log(q)$ range, and systematically  increases the values of \doha, especially in the tails. 

%%%%
%%   Subsec

\subsection{How do $\log(q)$ and \doha vary with galaxy properties?}\label{subsec:Results, GASP integrated functions}

To investigate if $\log(q)$ and \doha vary with integrated galactic properties, we compare galaxy global values (medians of spaxels) of $\log(q)$ and \doha as a function of stellar mass, SFR, specific SFR (sSFR=SFR/M$_\star$), and other overall galaxy properties. 
The integrated values of $\log(q)$ and \doha as a function of stellar mass are presented in Figures \ref{fig:Fig_stats_integrated} and \ref{fig:Fig_stats_integrated_p2}, respectively. 
The comparison with other properties (SFR and sSFR of disks, gas-phase metallicity, ratio of area of tails and disks) are shown in Appendix \ref{sec:Appendix 2}.

We separate panels and compare the median values of data as follows: 1) disks of control and stripped galaxy samples (panels A and C), 2) disks and tails of the stripped galaxies (panels B and D), and 3) dense gas (panels A and B) and DIG (C and D) dominated areas in the galaxies. 
We show the difference between DIG and dense gas data in panel E, and the difference between disks and tails of the stripped galaxies in panel F.  
From these comparisons, we observe a few major results and trends, as follows. 

The low-mass galaxies ($\rm \log_{10}M_{\ast}\leq10.3 \, M_{\odot}$) are characterized by a rather constant value of $\log(q)$, while $\log(q)$ for the high mass galaxies ($\rm \log_{10}M_{\ast}>10.3 \, M_{\odot}$) strongly correlates with stellar mass. 
DIG in disks and tails has $0.1-0.2$\,dex lower $\log(q)$ compared to the dense gas in the same environments. 
There is no clear difference in the evaluated gas properties in disks of control and stripped galaxies (panel E), and between the tails and disks of the stripped sample (panel F).
The large scatter in the measured properties prevents a conclusive analysis of trends with stellar mass in panels E and F.

Considering \doha of the dense gas, we observe a  weak anti-correlation with the stellar mass (Fig. \ref{fig:Fig_stats_integrated_p2}).
Furthermore, this anti-correlation is stronger for low mass galaxies than for high mass ones. 
The  DIG  areas show higher \doha values ($\approx 0.4$\,dex) compared to the dense gas (as clearly seen in panel E), with the tails having larger offsets (by $\approx0.1$\,dex) than the disks. 
We highlight  the fact that all dense gas data have \doha values below 0, while most of the DIG show positive values.  
We also notice in panel D that galaxies of higher J stage number have higher \doha values  for DIG designated spaxels in the tails (magenta square  symbols), compared to the rest of the data in other panels.   

Regarding  other galactic values (presented in Appendix \ref{sec:Appendix 2}), $\log(q)$ correlates with SFR and gas-phase metallicity. 
We explain this  as a signature of well known correlations between the  stellar mass, metallicity and SFR. 
The dense gas shows a somewhat strong anti-correlation between \doha and metallicity, and the DIG exhibits a weak correlation between \doha and  the spaxel fraction of tails vs. disks. 
The former relation is most likely  an effect of metallicity on increase of $\rm [O\textsc{I}]/H\alpha$ ratio. 

To conclude, these results indicate that the stellar mass is driving the increase in median $\log(q)$ across the galaxies more than the fraction of DIG. 
Moreover, the DIG clearly shows high \doha values compared to the dense gas, especially in the case of the stripped tails.

%%%%%%%%%%%%%%%%%%%%%%%%%%%
%%%%
%%   DISCUSSION
%%%%
%%%%%%%%%%%%%%%%%%%%%%%%%%%

\section{Summary of results and discussion}\label{sec: Discussion}

%%%%
%%   Subsec

\subsection{Variation with gas-phase metallicity}\label{subsec:Discussion, Z}

In the previous section we showed that the gas-phase metallicity at a given  galactocentric radius, and at kpc spatial resolution,  does not vary with different DIG fractions in disks of all galaxies (Fig. \ref{fig:Fig_diffZ}).
This would indicate two major conclusions about the distribution of the metallicity within the disks. 
First,  the DIG emission should not hinder the measurement of metallicity at a given radius in galaxies. 
Second, there is no difference in the metallicity radial gradient between areas dominated by   dense gas  or  DIG.

Similar to our conclusions, \citet{Kumari19}  did not found significant differences in metallicities between spatially close-by H\textsc{II}-DIG/LIER pairs, which are expected  to be chemically homogeneous.  
However, their results may be affected by non-SF source of ionisation changing  line ratios used for metallicity diagnostics.

On the other hand,  \citet{Poetrodjojo19} found tentative results  for different metallicity gradients in DIG and \hii dominated spaxels of the M83 galaxy.
Analysing IFU data from  MaNGA, \citet{Zhang17} also showed that the DIG can significantly bias the measurement of gas metallicity and metallicity gradients derived using strong emission lines. 
However, \citet{Zhang17} used an incomplete method to separate dense gas from DIG, by using only $\Sigma H\alpha$ information. 
\citet{Lacerda18} note that DIG in galaxy centers can exhibit bright $\Sigma H\alpha$ and low \wha, which could bias results of \citet{Zhang17}.  
Using data cubes from MaNGA and fiber spectroscopy from the Sloan Digital Sky Survey Data Release 7, \citet{Asari19} found that accounting for the DIG contribution leads to a change in metallicity of up to 0.1\,dex, but can be negligible depending on the metallicity estimation method.

The major drawback of the methods of separating DIG and \hii dominated spaxels,  used by  \citet{Poetrodjojo19} and \citet{Zhang17}, is that they did not fully account for effects of   radial metallicity gradients on \shar.
On the other hand, a drawback of our method is that the low spatial resolution potentially can dilute the metallicity values of the dense gas spaxels, as discussed by \citet{Sanchez14}, \citet{Belfiore17} and \citet{SanchezMeng2018}. This effect can lead to measuring similar metallicities between DIG and dense gas spaxels in our sample.

%%%%
%%   Subsec

\subsection{Ionization parameter}\label{subsec:Discussion, Q }

Our results at small spatial scales ($\approx1$ kpc) conclude that  the DIG exhibits lower $\log(q)$ than the dense gas  (Fig.  \ref{fig:Fig_diffZ} and  \ref{fig:Fig_stats_integrated}).
Similar conclusions have been drawn  by \citet{Zhang17} and \citet{Mingozzi20}  from their spatially resolved observations of $\log(q)$ in nearby galaxies.
This is to be expected, if we assume that the diffuse gas resides further from hot, young stars  that are the source of a large number of ionizing photons. 
Only exception at those spatial scales are DIG dominated spaxels in the edges of disks, and in the tails of  gas-stripped galaxies, which show  an increase in $\log(q)$.

Another interesting aspect of our results is that high-mass galaxies show an anti-correlation between  $\log(q)$ and galactocentric radius  (Fig. \ref{fig:Fig_diffZ}). 
We ascribe this increase to the ionization due to a large number of  HOLMES and evolved stars (\citealt{Flores-Fajardo2011}, \citealt{Zhang17})  at decreasing galactocentric radius.
Similarly, \citet{Dopita14}  observed higher $\log(q)$ values in the center of galaxies, and concluded that it is due to higher $\rm \Sigma SFR$ and changes in the geometry of the molecular and ionized gas. 
\citet{Simpson07} observed an increase in excitation of gas toward the galactic center, probably due to shocks.
On the other hand,  \citet{Mingozzi20} observed a correlation between $\log(q)$ and galactocentric radius, in high-mass galaxies. Similarly, \citet{Sanchez12} and \citet{Sanchez15} noted a weak correlation between those parameters.

There is an ongoing discussion about the relation between   the  ionizing parameter in galaxies and other integrated galactic properties. 
At galactic scales, we do not observe a strong effect of gas-stripping on  $\log(q)$, but we do note an increase in the $\log(q)$ range of the stripped galaxies compared to the control sample (Fig. \ref{fig:Fig_Galaxies int}).
On the other hand, this is the first study where a strong correlation between $\log(q)$ and the stellar mass for galaxies with $\log M_\ast/M_\odot>10.3 $ (Fig. \ref{fig:Fig_stats_integrated}) is observed. 
We ascribe this trend to higher fraction of HOLMES and evolved stars  within  high-mass galaxies, compared to low-mass galaxies.  
It is interesting that the tails also show an increase in $\log(q)$ with galaxy stellar mass, which may be a result of an  interplay between gas and photon densities of the ISM.  
Integrated SFR  and metallicities yield a positive correlation with $\log(q)$ probably due to their correlation with stellar mass.
Similar to our observations, \citet{Dopita14} observed a correlation between $\log(q)$ and SFR, while  \citet{Poetrodjojo18} did not.

%%%%
%%   Subsec

\subsection{Effects of gas-stripping process and \doha}\label{subsec:Discussion, DOHA and stripping}

Our results indicate an increase in \doha toward the edges of galaxy disks (Fig. \ref{fig:Fig_diffZ}), and a large  fraction of DIG dominated spaxels showing  positive \doha, designating them as LINER/LIER like regions (Fig. \ref{fig:Fig_Histo_spatial}). 
In general,  LINER/LIER like feature  for DIG dominated spaxels in centers of non-stripped galaxies  has been  previously observed by \citet{Flores-Fajardo2011} and  \citet{Zhang17}. 
\citet{Zhang17} speculate that only ionization by evolved stars (older than 125 Myr) as a major ionization source for DIG can lead to  LINER/LIER like emission. 
The evolved stars and HOLMES  may contribute to the increase in \doha within the disks of our sample (as seen in Fig. \ref{fig:Fig_diffZ} and \ref{fig:Fig_Histo_spatial}).

The DIG areas in the tails of  highly stripped galaxies (higher J stage numbers) systematically exhibit the highest \doha values compared to the galactic disks (Fig. \ref{fig:Fig_stats_integrated_p2}). 
\doha correlates with distance from the SF regions  in the disks and most of the tails (Fig. \ref{fig:Fig_Histo_spatial_dist}). 
We also found that  DIG in stripped tails extends to very large projected distances (Fig. \ref{fig:Fig_Histo_spatial_dist}), and is spanning kpc and up to 10 kpc scales, which is larger than a typical scale-heights of galactic disks.
The DIG at those large distances  contributes to the  increase  in \doha,  and  is showing  LINER/LIER like features. 
Those features are lacking in the   control galaxies.
Previously,  \citet{Poggianti19}  observed that the tails of many gas-stripped galaxies in GASP exhibit   LINER/LIER like emission in BPT-[O\textsc{i}] diagrams, while the other diagnostic diagrams (based on NII or SII) suggest star-formation or at least a composite origin for the ionization, which suggests that SF photons are partly the source  of ionization.   

In  light of  these results, we hypothesize that the DIG in the edges and tails of stripped galaxies,  is ionized by more than one process (of which star-formation may be one). 
Two competing  candidates for that ionization  are: 
1) older stellar population (older than 125 Myr and younger than 13 Gyr, as proposed by \citealt{Flores-Fajardo2011} and \citealt{Zhang17}) for ionization of DIG in galactic disks, or
2) mixing of the hot-cold gas layers such as ICM and galactic ISM, as proposed by  \citealt{Slavin93}, \citealt{Binette09}, and \citet{Poggianti19b}. 
The first option   is less likely because the older stellar population would most likely form in SF knots of tails and would cover small distances from those SF regions. 
Although this candidate of ionizing DIG in tails  should not be completely disregarded, we find  the mixing of hot-cold gas layers and shocks more probable.
This is also backed up by the findings of \citet{Poggianti19b}, \citealt{Campitiello21}  and \cite{Muller21}, who found an excess in X-ray emission and existence of magnetic fields in the tails of three of the jellyfish galaxies studied here, JO201, JW100, JO206.
This excess in X-ray may be produced by radiative cooling (\citealt{Muller21}, \citealt{Sparre20}) of the hot and highly magnetised ICM gas during the interplay with the galactic tail's ISM and accretion of the ICM onto the stripped ISM. 

To conclude, we hypothesize that the gas-stripping process is creating the galactic tails that are mixing with the ICM, thus ionizing the DIG at large distances from SF regions, and systematically  showing an  increase in \doha.

%%%%%%%%%%%%%%%%%%%%%%%%%%%
%%%%
%%   Conclusion
%%%%
%%%%%%%%%%%%%%%%%%%%%%%%%%%

\section{Conclusions}\label{sec: Conclusion}

In this paper, we utilise the optical IFU (MUSE spectrograph) observations of 71 galaxies (30 control and 41 stripped) from the GASP project  to analyze the gas properties of the dense and diffuse ionized gas.
The investigated properties  are gas-phase metallicity and ionization parameter $\log(q)$ for SF regions, and \doha values (the distance in the $\rm O\textsc{i}/H\alpha$  value from the BPT-$\rm [O\textsc{i}]$ line that separates SF and LINER/LIER like regions). 
We compare these properties  between  control sample  and stripped galaxies, as well as between the disks and tails of the stripped galaxies. 
Our results indicate the following: 

\begin{itemize}
    \item The DIG fraction does not strongly affect the gas-phase metallicity measurement at a given galactocentric radius, and thus  does not affect the metallicity radial gradient metallicity across all of the galaxies.
    
    \item The DIG dominated areas mostly show lower $\log(q)$  compared to the dense gas areas, at small spatial scales. 
    This is due to ionization of dense gas, in the disks and most of the tails, coming from close and highly-ionizing, young stars.
    
    \item  $\log(q)$ span a wider range in the stripped galaxies compared to the control sample, suggesting a larger spread in ionization parameter.  
    
    \item The integrated, galactic $\log(q)$ values correlate best with the total stellar mass, with  more massive galaxies showing also an increase in $\log(q)$ in their galactic centers. We speculate that this increase  is due to higher density of ionizing photons  from  older stellar populations  that are dominating the central regions,  and whose number is increasing at higher galaxy masses.

    \item The DIG systematically exhibits higher \doha compared to the dense gas, and all galaxies show an increase in \doha and LINER/LIER like features at their edges. 
    
    \item Gas-stripping process in galaxies create tails of gas, where mixing of the tail's ISM and ICM both  ionize the DIG, and cause an   increase in \doha values.   
    The areas of highest \doha reside in the tails of highly stripped galaxies (highest J stage number), in areas at large distances (up to 10 kpc) from the SF regions.
    These distances indicate that neither the SF regions alone,  nor the older stellar population alone or a combination of both, can explain the ionization of the DIG. 
    
\end{itemize}

These findings suggest that the gas-stripping process does not affect metallicities, nor has a strong effect on the ionization parameter, but has strong effects of gas-stripping  on \doha. 

Our results shed additional light on the process of ram-pressure that affects the ionized gas and star-formation in  galaxies. 
In the future, we aim to further investigate the dust and stellar ultraviolet emission of the DIG-dominated regions versus the dense regions, both in normal disks and disks and tails of stripped galaxies.

\acknowledgments

The authors wish to kindly thank Giovanni Fasano, who estimated positions and radii of H$\alpha$ clumps for GASP project. 
Based on observations collected at the European Organization for Astronomical Research in the Southern Hemisphere under ESO programme 196.B-0578. This project has received funding from the European Research Council (ERC) under the European Union's Horizon 2020 research and innovation programme (grant agreement No. 833824).
We acknowledge financial contribution from the contract ASI-INAF n.2017-14-H.0, from the grant PRIN MIUR 2017 n.20173ML3WW\_001 (PI Cimatti) and from the INAF main-stream funding programme (PI Vulcani). 
J.F. acknowledges financial support from the UNAM-DGAPA-PAPIIT IN111620 grant, Mexico.
This work  made use of Astropy, a community-developed core Python package for Astronomy  (\citealt{Astropy13}), and MPFIT (\citealt{MPFIT09}).

\vspace{5mm}
\facilities{MUSE, ESO}

\software{Astropy \citep{Astropy13}, KUBEVIZ \citep{Fossati16}, MPFIT \citep{Markwardt09}, PYQZ (v0.8.2; \citealt{Dopita13}, \citealt{Vogt15}),  SINOPSIS \citep{Fritz17} }

%%%%%%%%%%%%%%%%%%%%%%%%%%%
%%%%
%%   APPENDIX
%%%%
%%%%%%%%%%%%%%%%%%%%%%%%%%%

\clearpage

\appendix

 \section{Maps of all galaxies}
 \label{sec:Appendix 1}

\begin{figure*}[h!]
\centering
 \includegraphics[width=0.95\textwidth]{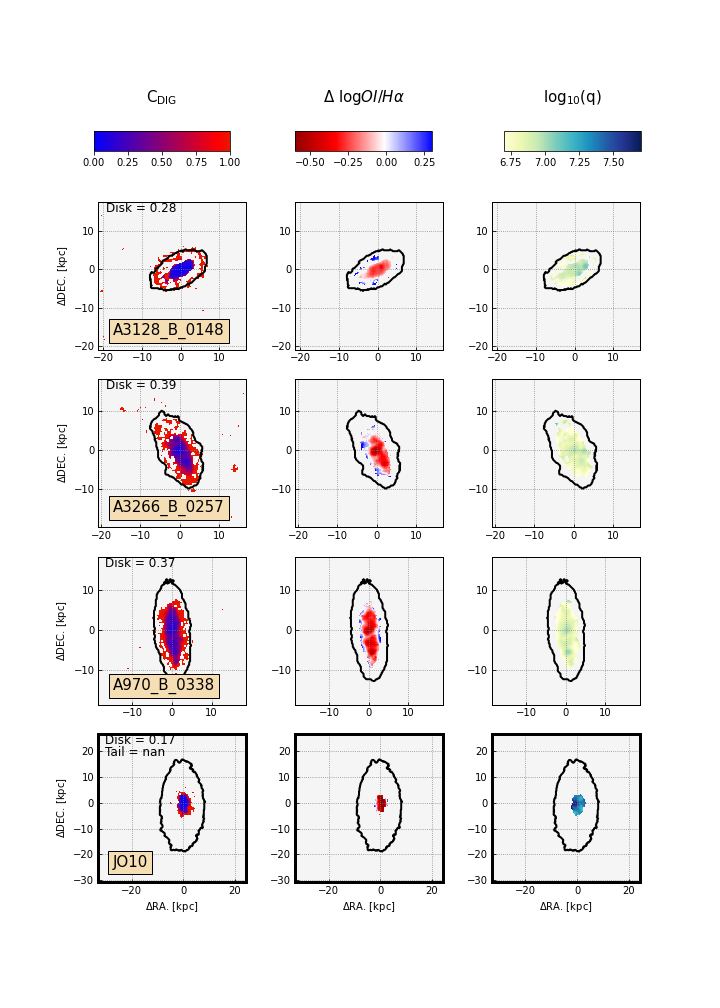}
   \caption{ Same as Fig. \ref{fig:Fig_Cdig_delOIHa_Q_p1}, but for all galaxies. Here, we put panels with control sample with thin edges,  and stripped galaxies  with  thick edges.     }
    \label{fig:App Cdig_maps v1}
\end{figure*}

\begin{figure*}[h!]
\centering
 \includegraphics[width=0.95\textwidth]{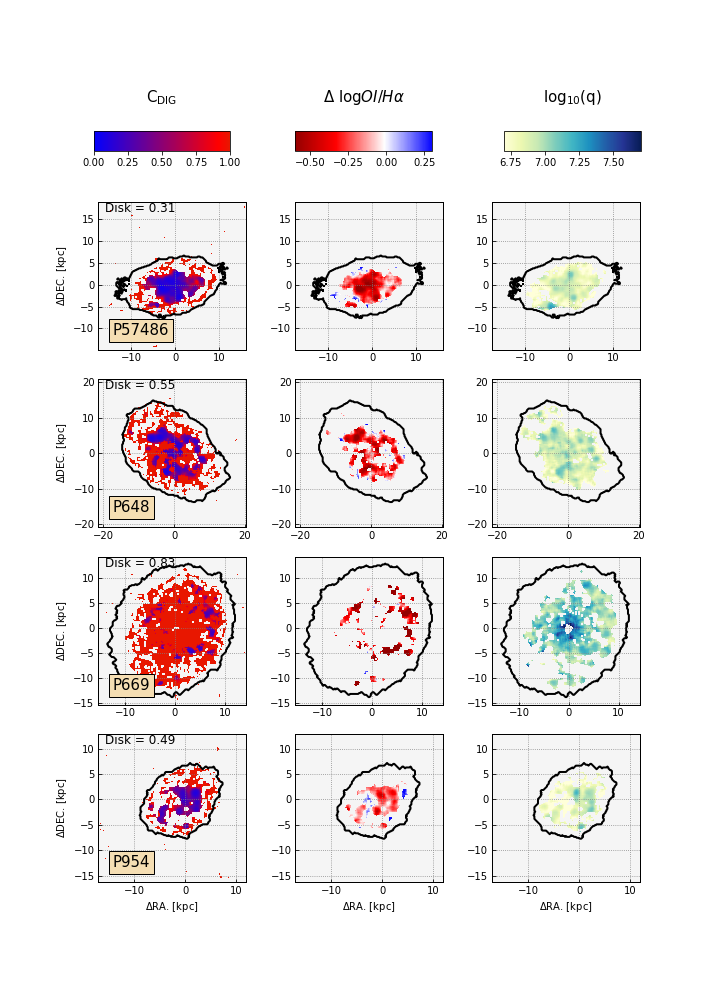}
   \caption{ Same as Fig. \ref{fig:Fig_Cdig_delOIHa_Q_p1}, but for all galaxies.   }
    \label{fig:App Cdig_maps v17}
\end{figure*}

Here (between Fig. \ref{fig:App Cdig_maps v1} and \ref{fig:App Cdig_maps v17}) we present $\rm C_{DIG}$, \doha , and  ionization parameter    maps of all the other galaxies in the sample, not shown in Fig. \ref{fig:Fig_Cdig_delOIHa_Q_p1}.
Here, we put panels with control sample and stripped galaxies with red or blue edges, respectively.

 \section{\doha and $\log(q)$ of integrated data}
 \label{sec:Appendix 2}

\begin{figure*}
\centering
\includegraphics[width=0.9\textwidth]{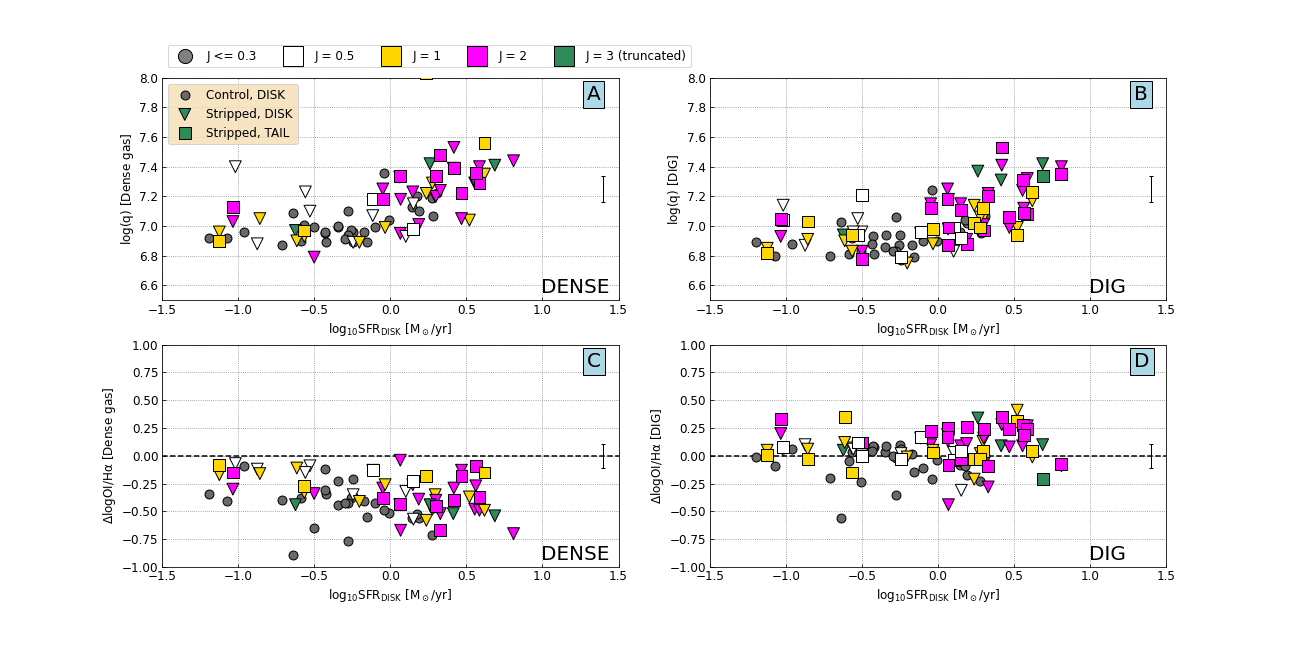}
\caption{  The median of  $\log(q)$ (panels A and B) and \doha (panels C and D) values for galaxies as a function of SFR. The galaxies are color-coded by J stage (labeled on the top), with only showing   galaxies with large enough number of spaxels in disks and tails. We present disks of the control and stripped samples, and tails of the stripped galaxies.  We  separated median of the dense gas dominated regions (panels A and C)  from medians from the DIG dominated areas (panels B and D). With the error-bar on the right, we show median of all error-bars for individual galaxies.    }
    \label{fig:App Int SFR}
\end{figure*}

\begin{figure*}
\centering
\includegraphics[width=0.9\textwidth]{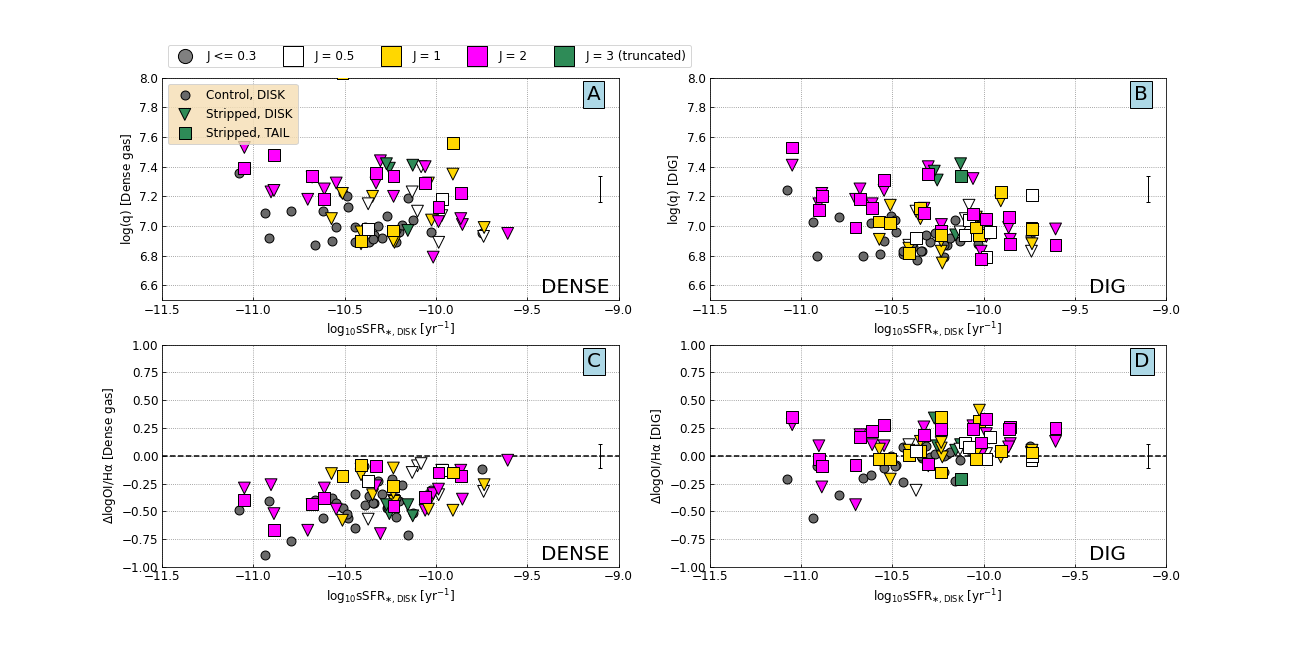}
\caption{  Same as Fig. \ref{fig:App Int SFR}, but on the x-axis is specific SFR (sSFR).    }
     \label{fig:App Int sSFR}
\end{figure*}

\begin{figure*}
\centering
\includegraphics[width=0.9\textwidth]{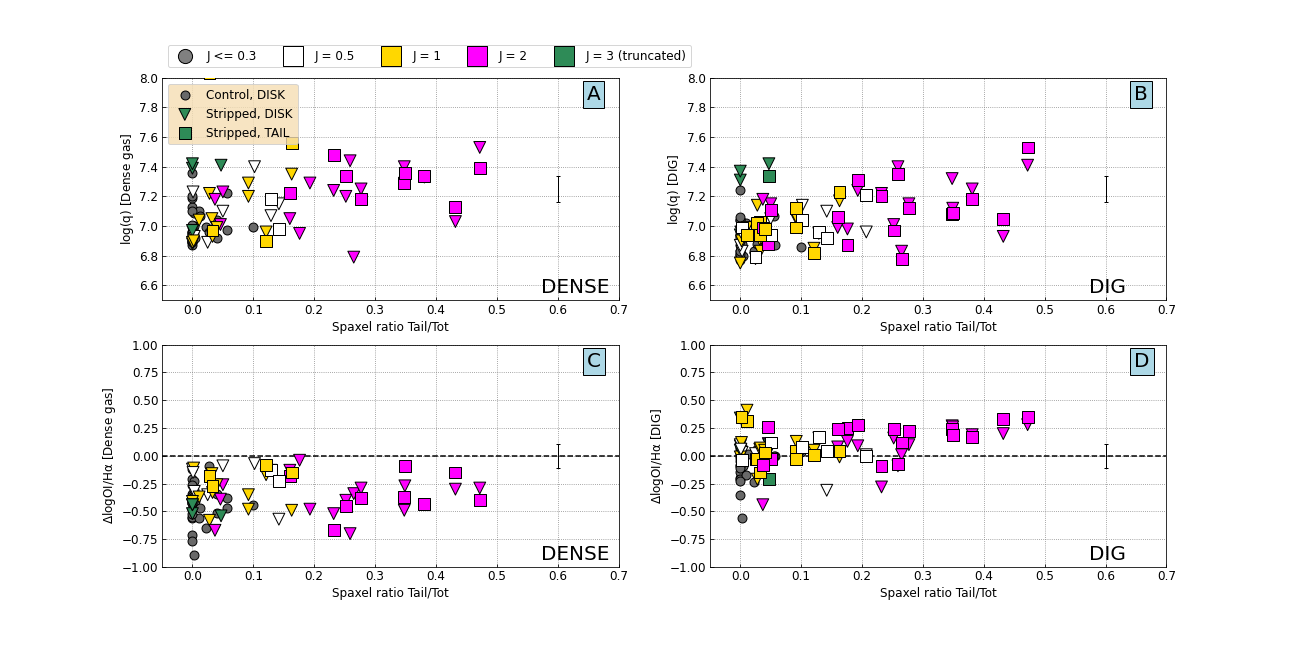}
\caption{  Same as Fig. \ref{fig:App Int SFR}, but on the x-axis is  ratio of numbers of spaxels from tail and disk.    }
     \label{fig:App Int PixRat}
\end{figure*}

\begin{figure*}
\centering
\includegraphics[width=0.9\textwidth]{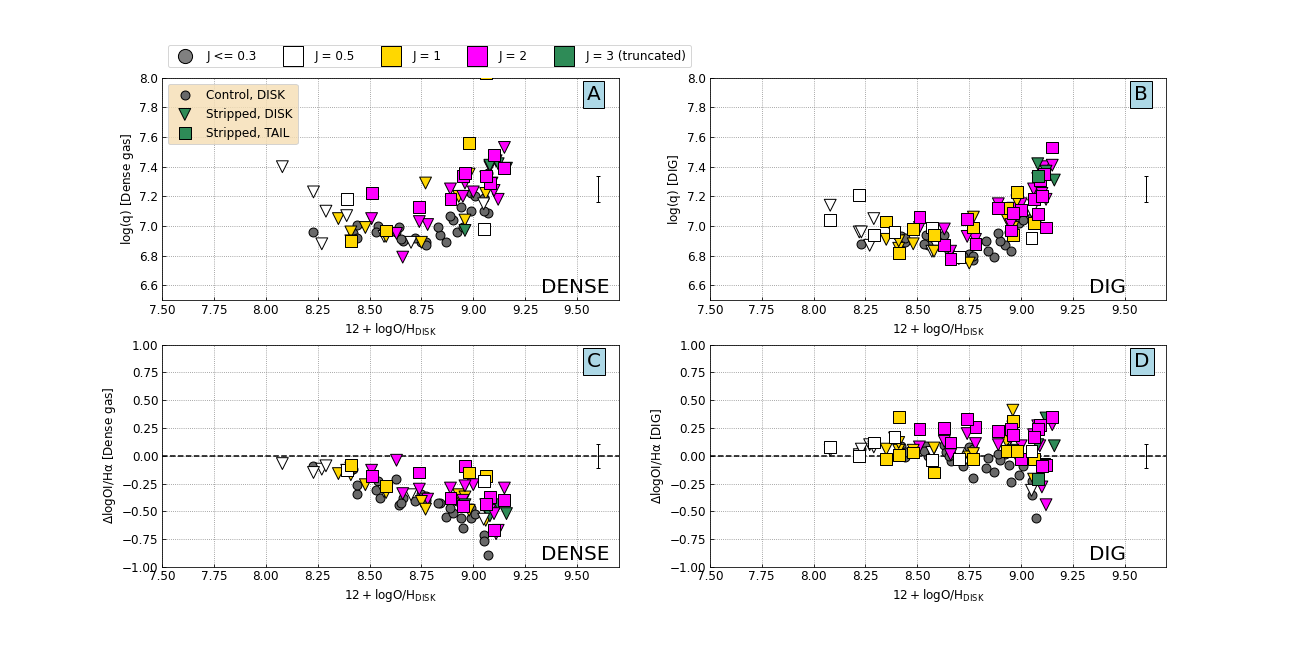}
\caption{  Same as Fig. \ref{fig:App Int SFR}, but on the x-axis is  gas-phase metallicity.    }
     \label{fig:App Int Z}
\end{figure*}

As in Fig. \ref{fig:Fig_stats_integrated} and \ref{fig:Fig_stats_integrated_p2}, in this section we compare integrated $\log(q)$ and \doha as a function of disk's SFR  (Fig. \ref{fig:App Int SFR}), sSFR (Fig. \ref{fig:App Int sSFR}), ratio of number of spaxels from tails and disks (Fig. \ref{fig:App Int PixRat}),  gas-phase metallicity (Fig. \ref{fig:App Int Z}). 
We separate the dense gas and DIG dominated data in these figures, and mark the data according to J stage (by color), or by which type or  part of galaxies it represent (type of marks). 

We hypothesize that the SFR and gas-phase metallicity  correlate to the mass of the stellar disk (\citealt{Franchetto20}), thus the behavior seen in the figures can be explained by variation in the mass instead of correlation with \doha and $\log(q)$. 
Similar correlations between integrated $\log(q)$ and SFR and metallicity was also  observed by \citet{Tremonti04} and  \citet{Dopita14}.
Furthermore, the metallicity values may be affected by $\log(q)$ measurements, so  trends seen in Fig. \ref{fig:App Int Z} may yield wrong  interpretation. 
For the sSFR and the spaxel number ratio of tails vs. disks, we do not see a clear correlation  in \doha and $\log(q)$.

\begin{figure*}[h!]
\centering
 \includegraphics[width=0.95\textwidth]{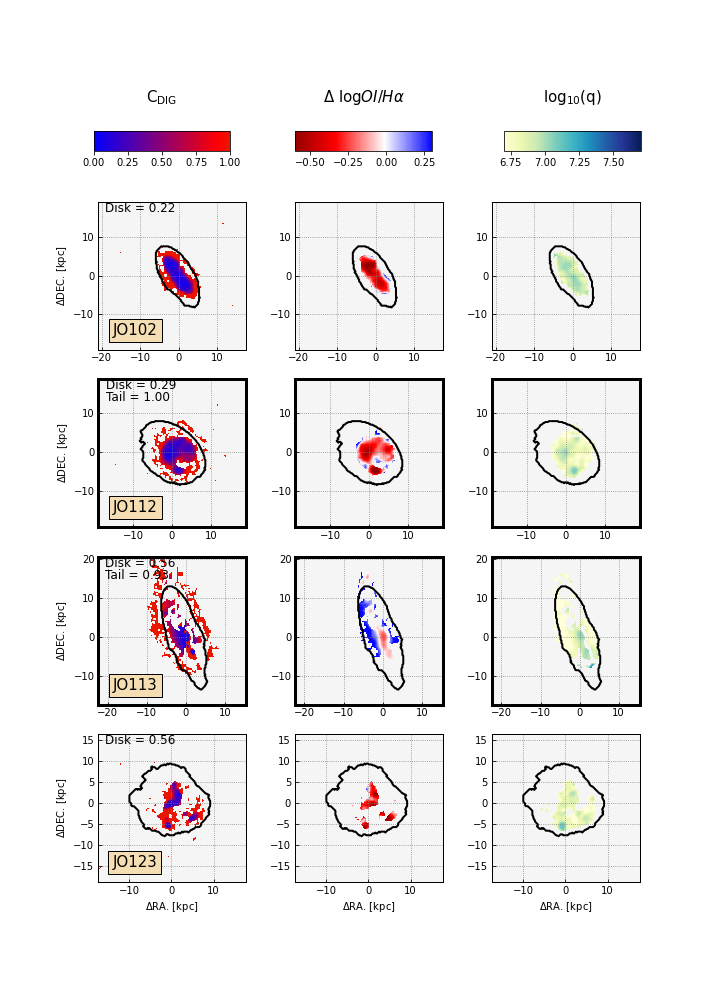}
   \caption{ Same as Fig. \ref{fig:Fig_Cdig_delOIHa_Q_p1}, but for all galaxies.   }
    \label{fig:App Cdig_maps v2}
\end{figure*}

\begin{figure*}[h!]
\centering
 \includegraphics[width=0.95\textwidth]{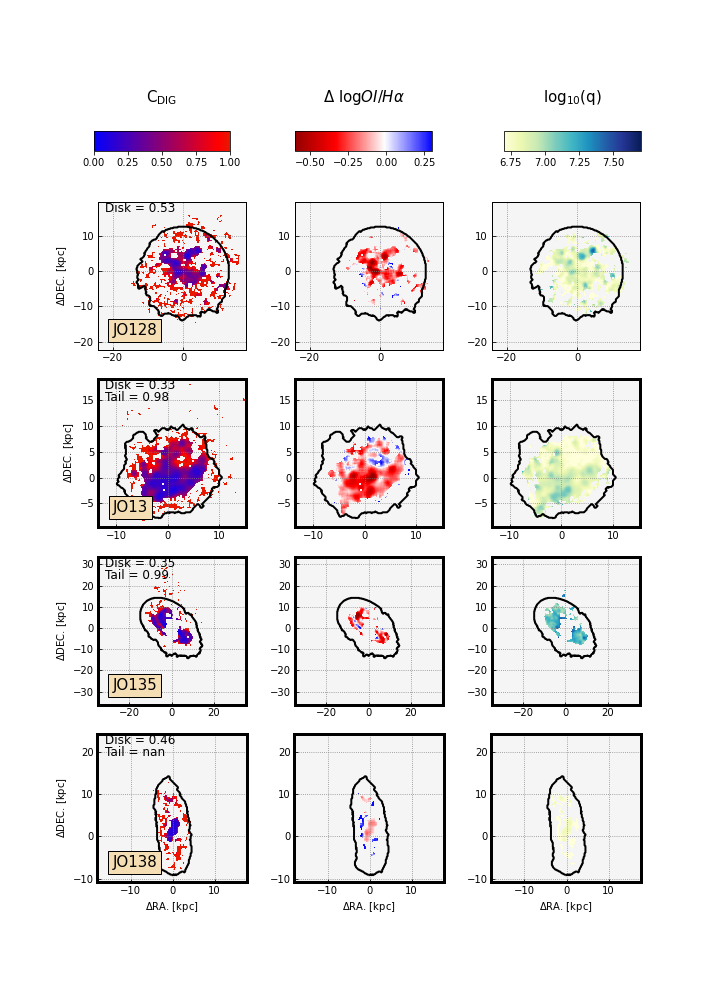}
   \caption{ Same as Fig. \ref{fig:Fig_Cdig_delOIHa_Q_p1}, but for all galaxies.   }
    \label{fig:App Cdig_maps v3}
\end{figure*}

\begin{figure*}[h!]
\centering
 \includegraphics[width=0.95\textwidth]{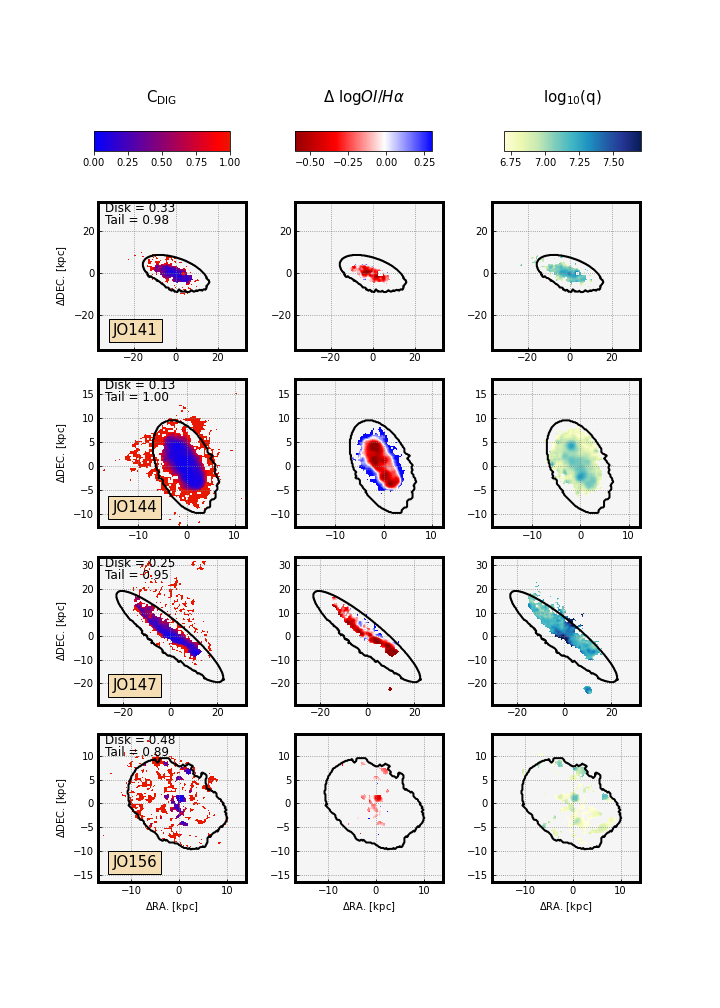}
   \caption{ Same as Fig. \ref{fig:Fig_Cdig_delOIHa_Q_p1}, but for all galaxies.   }
    \label{fig:App Cdig_maps v4}
\end{figure*}

\begin{figure*}[h!]
\centering
 \includegraphics[width=0.95\textwidth]{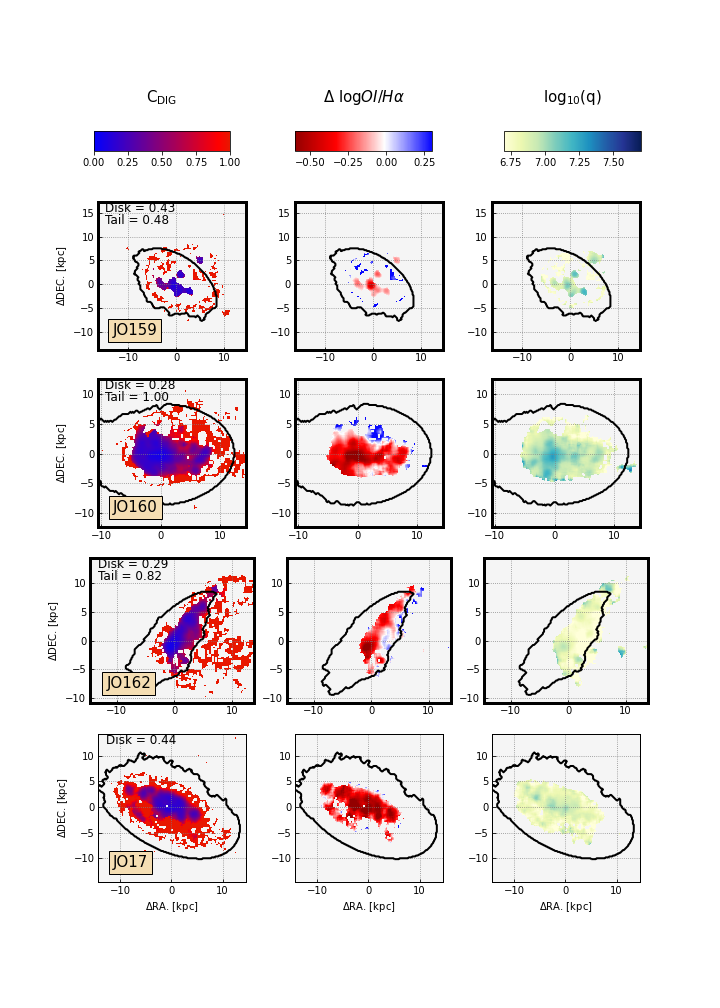}
   \caption{ Same as Fig. \ref{fig:Fig_Cdig_delOIHa_Q_p1}, but for all galaxies.   }
    \label{fig:App Cdig_maps v5}
\end{figure*}

\begin{figure*}[h!]
\centering
 \includegraphics[width=0.95\textwidth]{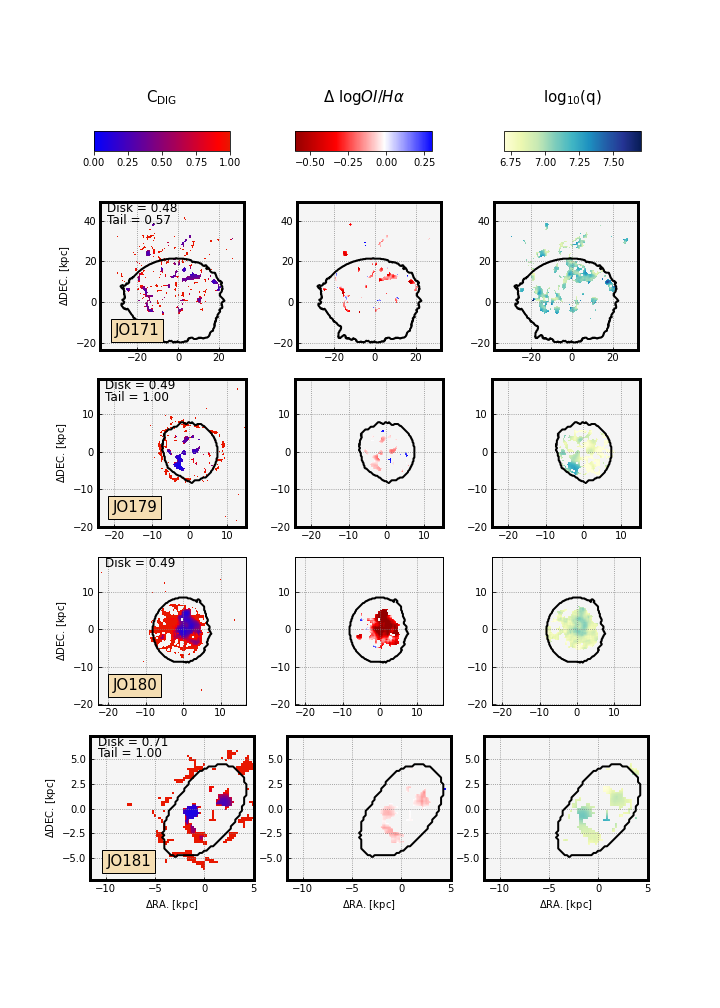}
   \caption{ Same as Fig. \ref{fig:Fig_Cdig_delOIHa_Q_p1}, but for all galaxies.   }
    \label{fig:App Cdig_maps v6}
\end{figure*}

\begin{figure*}[h!]
\centering
 \includegraphics[width=0.95\textwidth]{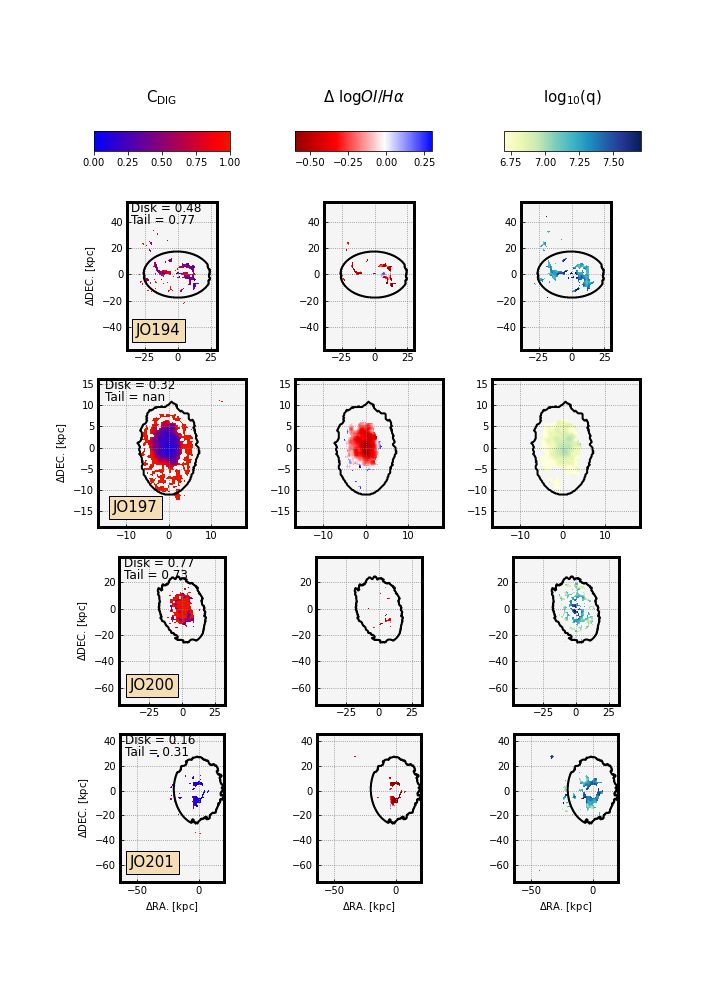}
   \caption{ Same as Fig. \ref{fig:Fig_Cdig_delOIHa_Q_p1}, but for all galaxies.   }
    \label{fig:App Cdig_maps v7}
\end{figure*}

\begin{figure*}[h!]
\centering
 \includegraphics[width=0.95\textwidth]{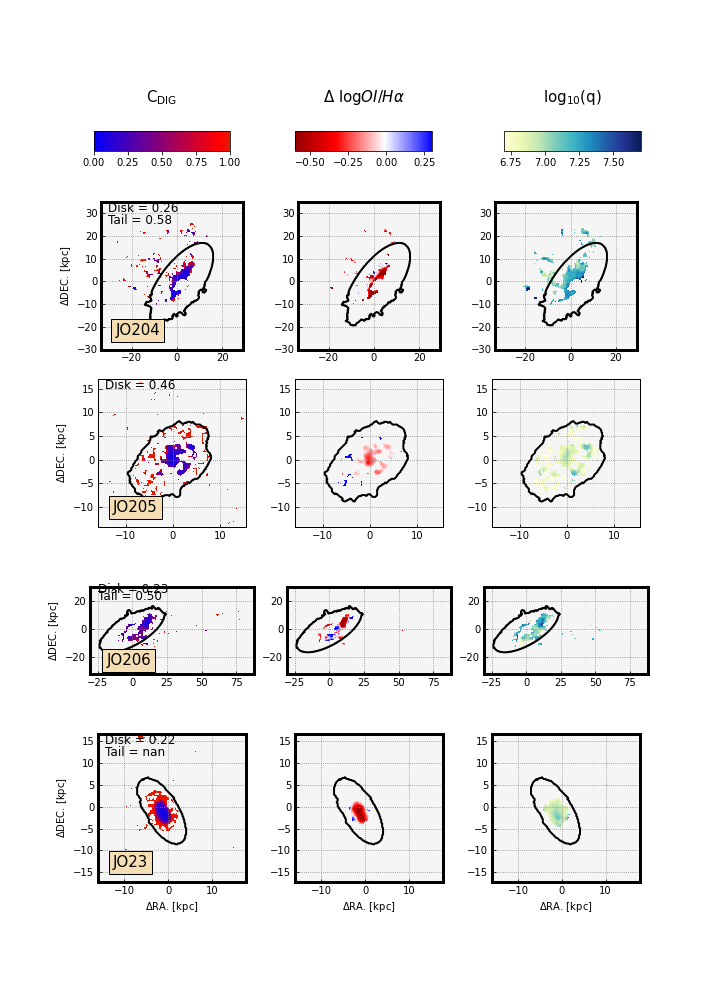}
   \caption{ Same as Fig. \ref{fig:Fig_Cdig_delOIHa_Q_p1}, but for all galaxies.   }
    \label{fig:App Cdig_maps v8}
\end{figure*}

\begin{figure*}[h!]
\centering
 \includegraphics[width=0.95\textwidth]{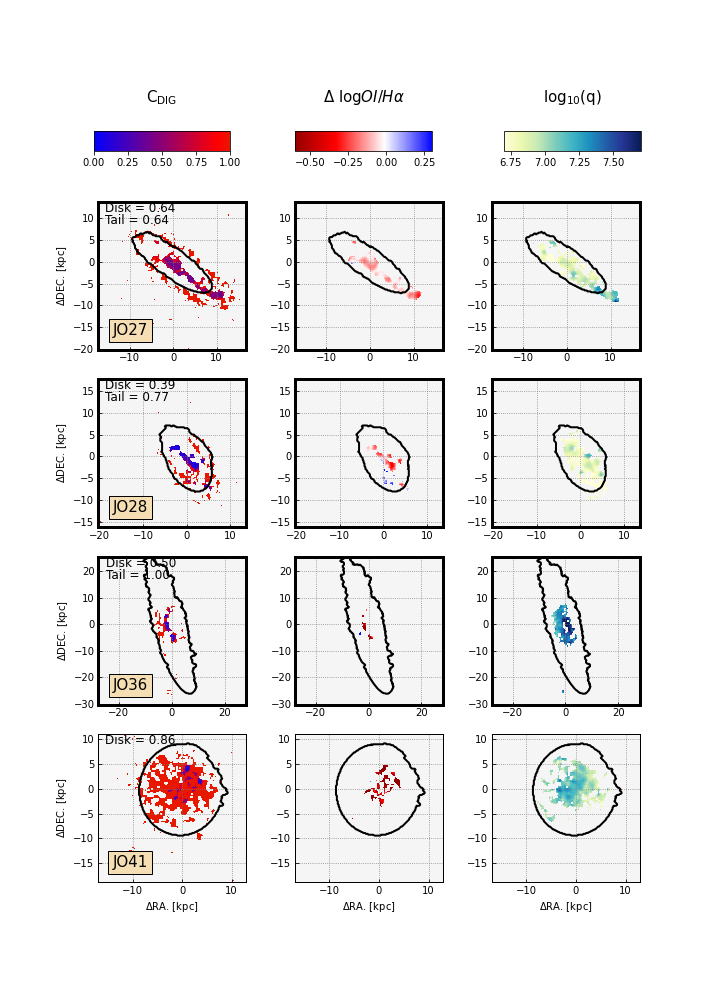}
   \caption{ Same as Fig. \ref{fig:Fig_Cdig_delOIHa_Q_p1}, but for all galaxies.   }
    \label{fig:App Cdig_maps v9}
\end{figure*}

\begin{figure*}[h!]
\centering
 \includegraphics[width=0.95\textwidth]{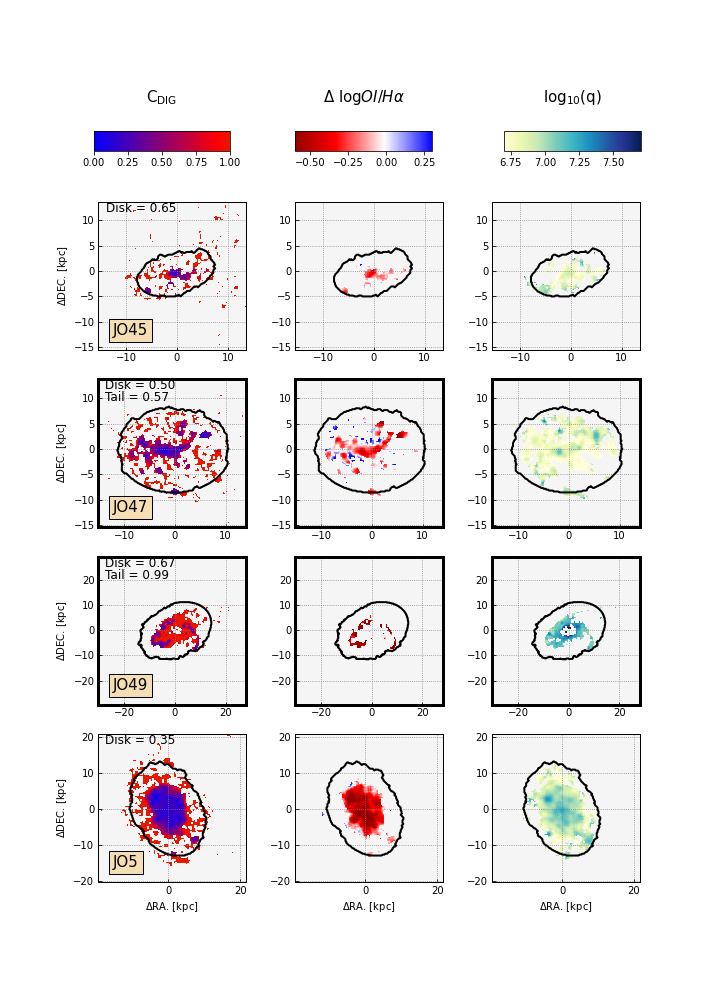}
   \caption{ Same as Fig. \ref{fig:Fig_Cdig_delOIHa_Q_p1}, but for all galaxies.   }
    \label{fig:App Cdig_maps v10}
\end{figure*}

\begin{figure*}[h!]
\centering
 \includegraphics[width=0.95\textwidth]{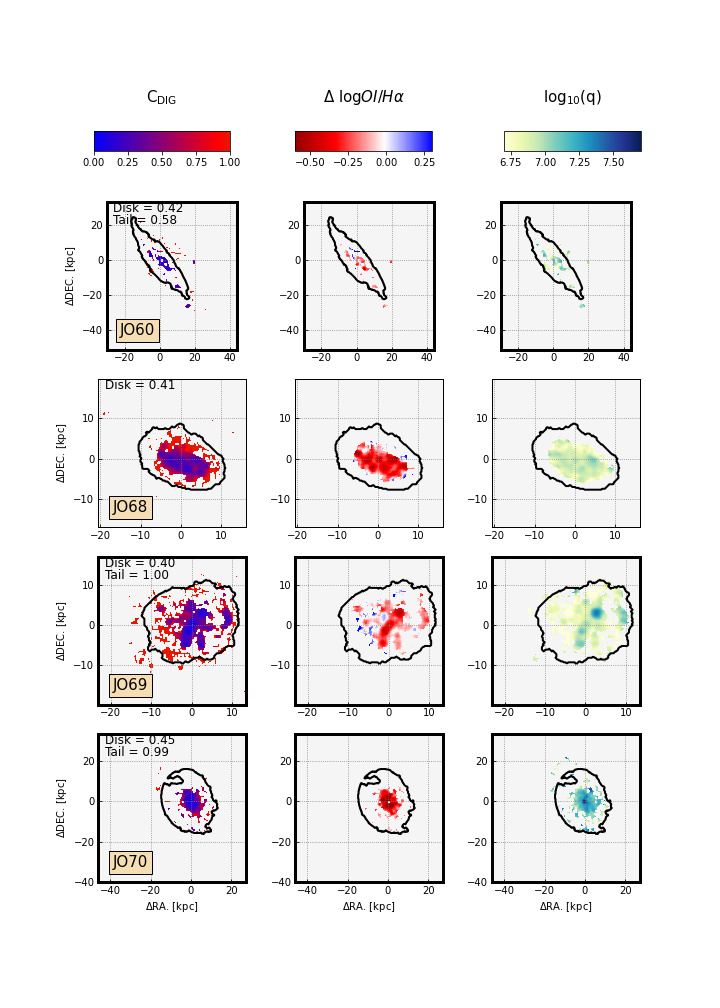}
   \caption{ Same as Fig. \ref{fig:Fig_Cdig_delOIHa_Q_p1}, but for all galaxies.   }
    \label{fig:App Cdig_maps v11}
\end{figure*}

\begin{figure*}[h!]
\centering
 \includegraphics[width=0.95\textwidth]{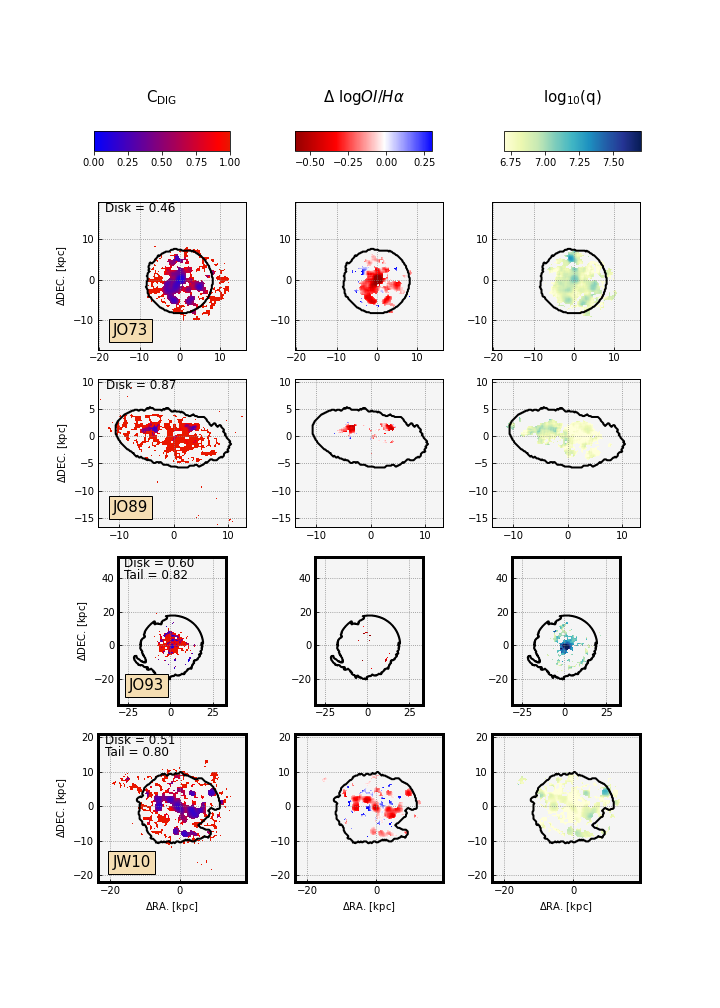}
   \caption{ Same as Fig. \ref{fig:Fig_Cdig_delOIHa_Q_p1}, but for all galaxies.   }
    \label{fig:App Cdig_maps v12}
\end{figure*}

\begin{figure*}[h!]
\centering
 \includegraphics[width=0.95\textwidth]{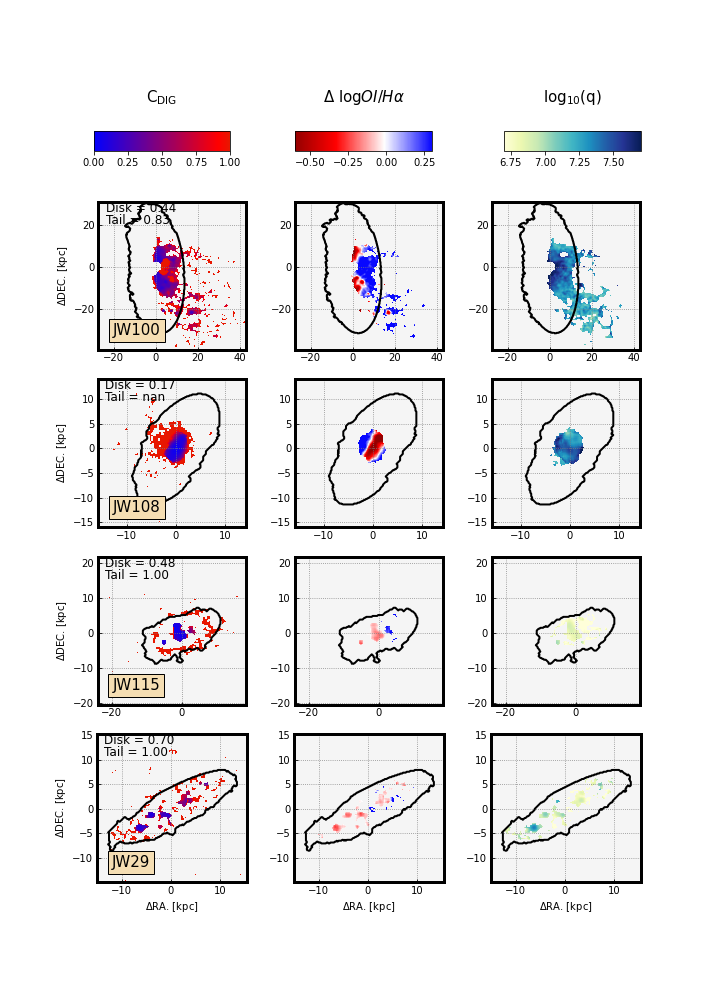}
   \caption{ Same as Fig. \ref{fig:Fig_Cdig_delOIHa_Q_p1}, but for all galaxies.   }
    \label{fig:App Cdig_maps v13}
\end{figure*}

\begin{figure*}[h!]
\centering
 \includegraphics[width=0.95\textwidth]{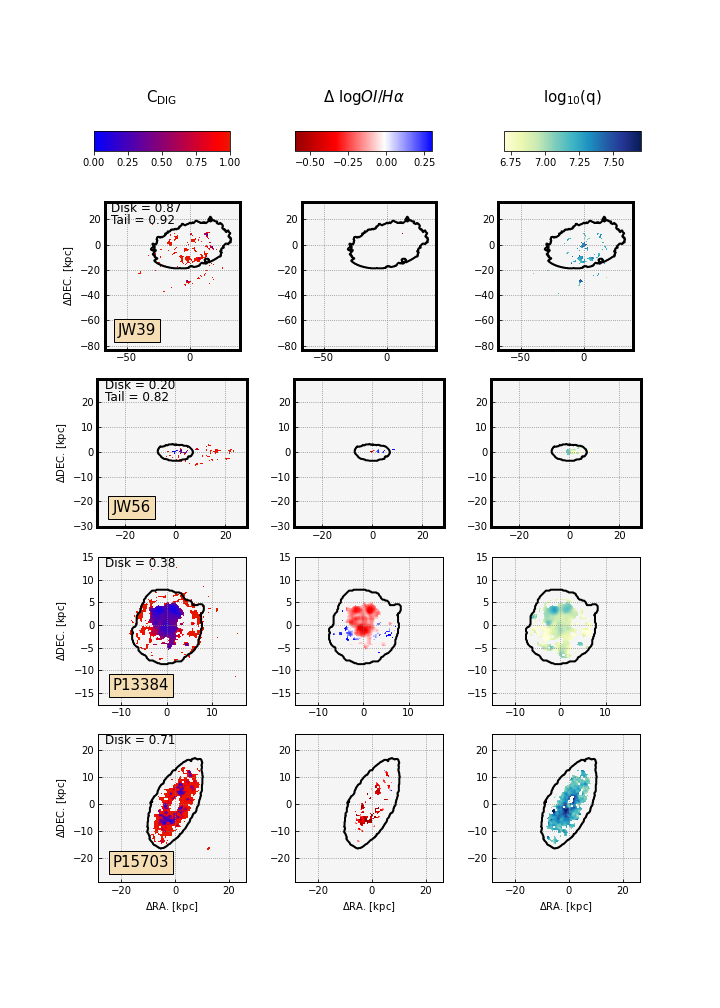}
   \caption{ Same as Fig. \ref{fig:Fig_Cdig_delOIHa_Q_p1}, but for all galaxies.   }
    \label{fig:App Cdig_maps v14}
\end{figure*}

\begin{figure*}[h!]
\centering
 \includegraphics[width=0.95\textwidth]{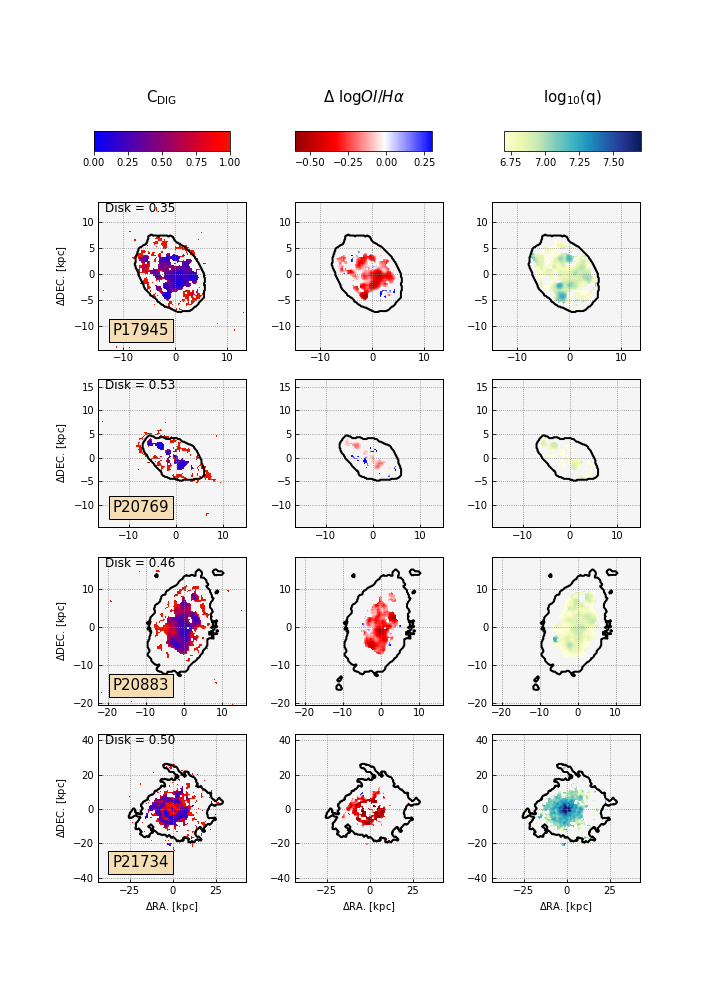}
   \caption{ Same as Fig. \ref{fig:Fig_Cdig_delOIHa_Q_p1}, but for all galaxies.   }
    \label{fig:App Cdig_maps v15}
\end{figure*}

\begin{figure*}[h!]
\centering
 \includegraphics[width=0.95\textwidth]{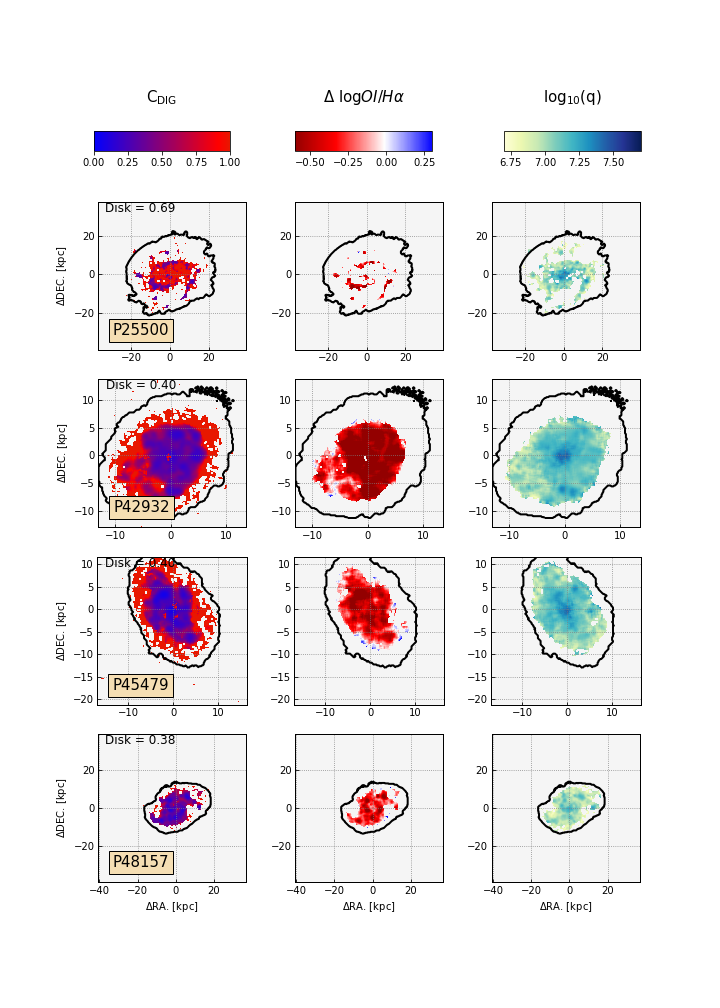}
   \caption{ Same as Fig. \ref{fig:Fig_Cdig_delOIHa_Q_p1}, but for all galaxies.   }
    \label{fig:App Cdig_maps v16}
\end{figure*}

%%%%%%%%%%%%%%%%%%%%%%%%%%%
%%%%
%%   BIBLIOGRAPHY
%%%%
%%%%%%%%%%%%%%%%%%%%%%%%%%%

\bibliographystyle{aasjournal}
\bibliography{NT_Citations}

\begin{thebibliography}{}
\expandafter\ifx\csname natexlab\endcsname\relax\def\natexlab#1{#1}\fi
\providecommand{\url}[1]{\href{#1}{#1}}
\providecommand{\dodoi}[1]{doi:~\href{http://doi.org/#1}{\nolinkurl{#1}}}
\providecommand{\doeprint}[1]{\href{http://ascl.net/#1}{\nolinkurl{http://ascl.net/#1}}}
\providecommand{\doarXiv}[1]{\href{https://arxiv.org/abs/#1}{\nolinkurl{https://arxiv.org/abs/#1}}}

\bibitem[{{Astropy Collaboration} {et~al.}(2013){Astropy Collaboration},
  {Robitaille}, {Tollerud}, {Greenfield}, {Droettboom}, {Bray}, {Aldcroft},
  {Davis}, {Ginsburg}, {Price-Whelan}, {Kerzendorf}, {Conley}, {Crighton},
  {Barbary}, {Muna}, {Ferguson}, {Grollier}, {Parikh}, {Nair}, {Unther},
  {Deil}, {Woillez}, {Conseil}, {Kramer}, {Turner}, {Singer}, {Fox}, {Weaver},
  {Zabalza}, {Edwards}, {Azalee Bostroem}, {Burke}, {Casey}, {Crawford},
  {Dencheva}, {Ely}, {Jenness}, {Labrie}, {Lim}, {Pierfederici}, {Pontzen},
  {Ptak}, {Refsdal}, {Servillat}, \& {Streicher}}]{Astropy13}
{Astropy Collaboration}, {Robitaille}, T.~P., {Tollerud}, E.~J., {et~al.} 2013,
  \aap, 558, A33, \dodoi{10.1051/0004-6361/201322068}

\bibitem[{Bacon {et~al.}(2006)Bacon, Bauer, Boehm, Boudon, Brau-Nogué,
  Caillier, Capoani, Carollo, Champavert, Contini, Daguisé, Dallé, Delabre,
  Devriendt, Dreizler, Dubois, Dupieux, Dupin, Emsellem, Ferruit, Franx,
  Gallou, Gerssen, Guiderdoni, Hahn, Hofmann, Jarno, Kelz, Koehler,
  Kollatschny, Kosmalski, Laurent, Lilly, Lizon, Loupias, Lynn, Manescau,
  McDermid, Monstein, Nicklas, Parès, Pasquini, Pécontal-Rousset, Pécontal,
  Pello, Petit, Picat, Popow, Quirrenbach, Reiss, Renault, Roth, Schaye,
  Soucail, Steinmetz, Stroebele, Stuik, Weilbacher, Wozniak, \&
  de~Zeeuw}]{Bacon10}
Bacon, R., Bauer, S., Boehm, P., {et~al.} 2006, in Ground-based and Airborne
  Instrumentation for Astronomy, ed. I.~S. McLean \& M.~Iye, Vol. 6269,
  International Society for Optics and Photonics (SPIE), 183 -- 191,
  \dodoi{10.1117/12.669772}

\bibitem[{{Baldwin} {et~al.}(1981){Baldwin}, {Phillips}, \&
  {Terlevich}}]{Baldwin81}
{Baldwin}, J.~A., {Phillips}, M.~M., \& {Terlevich}, R. 1981, \pasp, 93, 5,
  \dodoi{10.1086/130766}

\bibitem[{{Barnes} {et~al.}(2014){Barnes}, {Wood}, {Hill}, \&
  {Haffner}}]{Barnes14}
{Barnes}, J.~E., {Wood}, K., {Hill}, A.~S., \& {Haffner}, L.~M. 2014, \mnras,
  440, 3027, \dodoi{10.1093/mnras/stu521}

\bibitem[{{Belfiore} {et~al.}(2016){Belfiore}, {Maiolino}, {Maraston},
  {Emsellem}, {Bershady}, {Masters}, {Yan}, {Bizyaev}, {Boquien}, {Brownstein},
  {Bundy}, {Drory}, {Heckman}, {Law}, {Roman-Lopes}, {Pan}, {Stanghellini},
  {Thomas}, {Weijmans}, \& {Westfall}}]{Belfiore16}
{Belfiore}, F., {Maiolino}, R., {Maraston}, C., {et~al.} 2016, \mnras, 461,
  3111, \dodoi{10.1093/mnras/stw1234}

\bibitem[{{Belfiore} {et~al.}(2017{\natexlab{a}}){Belfiore}, {Maiolino},
  {Maraston}, {Emsellem}, {Bershady}, {Masters}, {Bizyaev}, {Boquien},
  {Brownstein}, {Bundy}, {Diamond-Stanic}, {Drory}, {Heckman}, {Law},
  {Malanushenko}, {Oravetz}, {Pan}, {Roman-Lopes}, {Thomas}, {Weijmans},
  {Westfall}, \& {Yan}}]{Belfiore17b}
---. 2017{\natexlab{a}}, \mnras, 466, 2570, \dodoi{10.1093/mnras/stw3211}

\bibitem[{{Belfiore} {et~al.}(2017{\natexlab{b}}){Belfiore}, {Maiolino},
  {Tremonti}, {S{\'a}nchez}, {Bundy}, {Bershady}, {Westfall}, {Lin}, {Drory},
  {Boquien}, {Thomas}, \& {Brinkmann}}]{Belfiore17}
{Belfiore}, F., {Maiolino}, R., {Tremonti}, C., {et~al.} 2017{\natexlab{b}},
  \mnras, 469, 151, \dodoi{10.1093/mnras/stx789}

\bibitem[{{Binette} {et~al.}(2009){Binette}, {Drissen}, {Ubeda}, {Raga},
  {Robert}, \& {Krongold}}]{Binette09}
{Binette}, L., {Drissen}, L., {Ubeda}, L., {et~al.} 2009, \aap, 500, 817,
  \dodoi{10.1051/0004-6361/200811132}

\bibitem[{{Blanc} {et~al.}(2009){Blanc}, {Heiderman}, {Gebhardt}, {Evans}, \&
  {Adams}}]{Blanc09}
{Blanc}, G.~A., {Heiderman}, A., {Gebhardt}, K., {Evans}, Neal~J., I., \&
  {Adams}, J. 2009, \apj, 704, 842, \dodoi{10.1088/0004-637X/704/1/842}

\bibitem[{{Bocchio} {et~al.}(2016){Bocchio}, {Bianchi}, {Hunt}, \&
  {Schneider}}]{Bocchio16}
{Bocchio}, M., {Bianchi}, S., {Hunt}, L.~K., \& {Schneider}, R. 2016, \aap,
  586, A8, \dodoi{10.1051/0004-6361/201526950}

\bibitem[{{Bundy} {et~al.}(2015){Bundy}, {Bershady}, {Law}, {Yan}, {Drory},
  {MacDonald}, {Wake}, {Cherinka}, {S{\'a}nchez-Gallego}, {Weijmans}, {Thomas},
  {Tremonti}, {Masters}, {Coccato}, {Diamond-Stanic}, {Arag{\'o}n-Salamanca},
  {Avila-Reese}, {Badenes}, {Falc{\'o}n-Barroso}, {Belfiore}, {Bizyaev},
  {Blanc}, {Bland-Hawthorn}, {Blanton}, {Brownstein}, {Byler}, {Cappellari},
  {Conroy}, {Dutton}, {Emsellem}, {Etherington}, {Frinchaboy}, {Fu}, {Gunn},
  {Harding}, {Johnston}, {Kauffmann}, {Kinemuchi}, {Klaene}, {Knapen},
  {Leauthaud}, {Li}, {Lin}, {Maiolino}, {Malanushenko}, {Malanushenko}, {Mao},
  {Maraston}, {McDermid}, {Merrifield}, {Nichol}, {Oravetz}, {Pan}, {Parejko},
  {Sanchez}, {Schlegel}, {Simmons}, {Steele}, {Steinmetz}, {Thanjavur},
  {Thompson}, {Tinker}, {van den Bosch}, {Westfall}, {Wilkinson}, {Wright},
  {Xiao}, \& {Zhang}}]{Bundy15}
{Bundy}, K., {Bershady}, M.~A., {Law}, D.~R., {et~al.} 2015, \apj, 798, 7,
  \dodoi{10.1088/0004-637X/798/1/7}

\bibitem[{{Calvi} {et~al.}(2011){Calvi}, {Poggianti}, \& {Vulcani}}]{Calvi11}
{Calvi}, R., {Poggianti}, B.~M., \& {Vulcani}, B. 2011, \mnras, 416, 727,
  \dodoi{10.1111/j.1365-2966.2011.19088.x}

\bibitem[{{Campitiello} {et~al.}(2021){Campitiello}, {Ignesti}, {Gitti},
  {Brighenti}, {Radovich}, {Wolter}, {Tomicic}, {Bellhouse}, {Poggianti},
  {Moretti}, {Vulcani}, {Jaff{\`e}}, {Paladino}, {Muller}, {Fritz}, {Lourenco},
  \& {Gullieuszik}}]{Campitiello21}
{Campitiello}, M.~G., {Ignesti}, A., {Gitti}, M., {et~al.} 2021, arXiv
  e-prints, arXiv:2103.03147.
\newblock \doarXiv{2103.03147}

\bibitem[{{Cardelli} {et~al.}(1989){Cardelli}, {Clayton}, \&
  {Mathis}}]{Cardelli89}
{Cardelli}, J.~A., {Clayton}, G.~C., \& {Mathis}, J.~S. 1989, \apj, 345, 245,
  \dodoi{10.1086/167900}

\bibitem[{{Chabrier}(2003)}]{Chabrier03}
{Chabrier}, G. 2003, \pasp, 115, 763, \dodoi{10.1086/376392}

\bibitem[{{Cid Fernandes} {et~al.}(2011){Cid Fernandes}, {Stasi{\'n}ska},
  {Mateus}, \& {Vale Asari}}]{Cid11}
{Cid Fernandes}, R., {Stasi{\'n}ska}, G., {Mateus}, A., \& {Vale Asari}, N.
  2011, \mnras, 413, 1687, \dodoi{10.1111/j.1365-2966.2011.18244.x}

\bibitem[{{Collins} \& {Rand}(2001)}]{Collins01}
{Collins}, J.~A., \& {Rand}, R.~J. 2001, \apj, 551, 57, \dodoi{10.1086/320072}

\bibitem[{{Consolandi} {et~al.}(2017){Consolandi}, {Gavazzi}, {Fossati},
  {Fumagalli}, {Boselli}, {Yagi}, \& {Yoshida}}]{Consolandi17}
{Consolandi}, G., {Gavazzi}, G., {Fossati}, M., {et~al.} 2017, \aap, 606, A83,
  \dodoi{10.1051/0004-6361/201731218}

\bibitem[{{Cowie} \& {Songaila}(1977)}]{CowieSongaila77}
{Cowie}, L.~L., \& {Songaila}, A. 1977, \nat, 266, 501,
  \dodoi{10.1038/266501a0}

\bibitem[{{Della Bruna} {et~al.}(2020){Della Bruna}, {Adamo}, {Bik},
  {Fumagalli}, {Walterbos}, {{\"O}stlin}, {Bruzual}, {Calzetti}, {Charlot},
  {Grasha}, {Smith}, {Thilker}, \& {Wofford}}]{Bruna20}
{Della Bruna}, L., {Adamo}, A., {Bik}, A., {et~al.} 2020, \aap, 635, A134,
  \dodoi{10.1051/0004-6361/201937173}

\bibitem[{{Dopita} {et~al.}(2014){Dopita}, {Rich}, {Vogt}, {Kewley}, {Ho},
  {Basurah}, {Ali}, \& {Amer}}]{Dopita14}
{Dopita}, M.~A., {Rich}, J., {Vogt}, F. P.~A., {et~al.} 2014, \apss, 350, 741,
  \dodoi{10.1007/s10509-013-1753-2}

\bibitem[{{Dopita} {et~al.}(2013){Dopita}, {Sutherland}, {Nicholls}, {Kewley},
  \& {Vogt}}]{Dopita13}
{Dopita}, M.~A., {Sutherland}, R.~S., {Nicholls}, D.~C., {Kewley}, L.~J., \&
  {Vogt}, F. P.~A. 2013, \apjs, 208, 10, \dodoi{10.1088/0067-0049/208/1/10}

\bibitem[{{Fasano} {et~al.}(2006){Fasano}, {Marmo}, {Varela}, {D'Onofrio},
  {Poggianti}, {Moles}, {Pignatelli}, {Bettoni}, {Kj{\ae}rgaard}, {Rizzi},
  {Couch}, \& {Dressler}}]{Fasano06}
{Fasano}, G., {Marmo}, C., {Varela}, J., {et~al.} 2006, \aap, 445, 805,
  \dodoi{10.1051/0004-6361:20053816}

\bibitem[{{Flores-Fajardo} {et~al.}(2011){Flores-Fajardo}, {Morisset},
  {Stasi{\'n}ska}, \& {Binette}}]{Flores-Fajardo2011}
{Flores-Fajardo}, N., {Morisset}, C., {Stasi{\'n}ska}, G., \& {Binette}, L.
  2011, \mnras, 415, 2182, \dodoi{10.1111/j.1365-2966.2011.18848.x}

\bibitem[{{Fossati} {et~al.}(2016){Fossati}, {Fumagalli}, {Boselli}, {Gavazzi},
  {Sun}, \& {Wilman}}]{Fossati16}
{Fossati}, M., {Fumagalli}, M., {Boselli}, A., {et~al.} 2016, \mnras, 455,
  2028, \dodoi{10.1093/mnras/stv2400}

\bibitem[{{Franchetto} {et~al.}(2020){Franchetto}, {Vulcani}, {Poggianti},
  {Gullieuszik}, {Mingozzi}, {Moretti}, {Tomi{\v{c}}i{\'c}}, {Fritz},
  {Bettoni}, \& {Jaff{\'e}}}]{Franchetto20}
{Franchetto}, A., {Vulcani}, B., {Poggianti}, B.~M., {et~al.} 2020, \apj, 895,
  106, \dodoi{10.3847/1538-4357/ab8db9}

\bibitem[{{Fritz} {et~al.}(2017){Fritz}, {Moretti}, {Gullieuszik}, {Poggianti},
  {Bruzual}, {Vulcani}, {Nicastro}, {Jaff{\'e}}, {Cervantes Sodi}, {Bettoni},
  {Biviano}, {Fasano}, {Charlot}, {Bellhouse}, \& {Hau}}]{Fritz17}
{Fritz}, J., {Moretti}, A., {Gullieuszik}, M., {et~al.} 2017, \apj, 848, 132,
  \dodoi{10.3847/1538-4357/aa8f51}

\bibitem[{{Fumagalli} {et~al.}(2014){Fumagalli}, {Fossati}, {Hau}, {Gavazzi},
  {Bower}, {Sun}, \& {Boselli}}]{Fumagalli14}
{Fumagalli}, M., {Fossati}, M., {Hau}, G. K.~T., {et~al.} 2014, \mnras, 445,
  4335, \dodoi{10.1093/mnras/stu2092}

\bibitem[{{Gullieuszik} {et~al.}(2015){Gullieuszik}, {Poggianti}, {Fasano},
  {Zaggia}, {Paccagnella}, {Moretti}, {Bettoni}, {D'Onofrio}, {Couch},
  {Vulcani}, {Fritz}, {Omizzolo}, {Baruffolo}, {Schipani}, {Capaccioli}, \&
  {Varela}}]{Gullieuszik15}
{Gullieuszik}, M., {Poggianti}, B., {Fasano}, G., {et~al.} 2015, \aap, 581,
  A41, \dodoi{10.1051/0004-6361/201526061}

\bibitem[{{Gullieuszik} {et~al.}(2020){Gullieuszik}, {Poggianti}, {McGee},
  {Moretti}, {Vulcani}, {Tonnesen}, {Roediger}, {Jaff{\'e}}, {Fritz},
  {Franchetto}, {Omizzolo}, {Bettoni}, {Radovich}, \& {Wolter}}]{Gullieuszik20}
{Gullieuszik}, M., {Poggianti}, B.~M., {McGee}, S.~L., {et~al.} 2020, arXiv
  e-prints, arXiv:2006.16032.
\newblock \doarXiv{2006.16032}

\bibitem[{{Gunn} \& {Gott}(1972)}]{GunnGott72}
{Gunn}, J.~E., \& {Gott}, J.~Richard, I. 1972, \apj, 176, 1,
  \dodoi{10.1086/151605}

\bibitem[{{Haffner} {et~al.}(2009){Haffner}, {Dettmar}, {Beckman}, {Wood},
  {Slavin}, {Giammanco}, {Madsen}, {Zurita}, \& {Reynolds}}]{Haffner09}
{Haffner}, L.~M., {Dettmar}, R.~J., {Beckman}, J.~E., {et~al.} 2009, Reviews of
  Modern Physics, 81, 969, \dodoi{10.1103/RevModPhys.81.969}

\bibitem[{{Heckman} \& {Balick}(1980)}]{Heckman80}
{Heckman}, T.~M., \& {Balick}, B. 1980, \aap, 83, 100

\bibitem[{{Herpich} {et~al.}(2018){Herpich}, {Stasi{\'n}ska}, {Mateus}, {Vale
  Asari}, \& {Cid Fernandes}}]{Herpich18}
{Herpich}, F., {Stasi{\'n}ska}, G., {Mateus}, A., {Vale Asari}, N., \& {Cid
  Fernandes}, R. 2018, \mnras, 481, 1774, \dodoi{10.1093/mnras/sty2391}

\bibitem[{{Hoopes} \& {Walterbos}(2003)}]{Hoopes03}
{Hoopes}, C.~G., \& {Walterbos}, R. A.~M. 2003, \apj, 586, 902,
  \dodoi{10.1086/367954}

\bibitem[{{Hummer} \& {Storey}(1987)}]{HummerStorey87}
{Hummer}, D.~G., \& {Storey}, P.~J. 1987, \mnras, 224, 801,
  \dodoi{10.1093/mnras/224.3.801}

\bibitem[{{Kaplan} {et~al.}(2016){Kaplan}, {Jogee}, {Kewley}, {Blanc},
  {Weinzirl}, {Song}, {Drory}, {Luo}, \& {van den Bosch}}]{Kaplan16}
{Kaplan}, K.~F., {Jogee}, S., {Kewley}, L., {et~al.} 2016, \mnras, 462, 1642,
  \dodoi{10.1093/mnras/stw1422}

\bibitem[{{Kewley} {et~al.}(2001){Kewley}, {Dopita}, {Sutherland}, {Heisler},
  \& {Trevena}}]{Kewley01}
{Kewley}, L.~J., {Dopita}, M.~A., {Sutherland}, R.~S., {Heisler}, C.~A., \&
  {Trevena}, J. 2001, \apj, 556, 121, \dodoi{10.1086/321545}

\bibitem[{{Kewley} {et~al.}(2006){Kewley}, {Groves}, {Kauffmann}, \&
  {Heckman}}]{Kewley06}
{Kewley}, L.~J., {Groves}, B., {Kauffmann}, G., \& {Heckman}, T. 2006, \mnras,
  372, 961, \dodoi{10.1111/j.1365-2966.2006.10859.x}

\bibitem[{{Kewley} {et~al.}(2019){Kewley}, {Nicholls}, \&
  {Sutherland}}]{Kewley19}
{Kewley}, L.~J., {Nicholls}, D.~C., \& {Sutherland}, R.~S. 2019, \araa, 57,
  511, \dodoi{10.1146/annurev-astro-081817-051832}

\bibitem[{Koopmann \& Kenney(2004)}]{Koopmann04}
Koopmann, R.~A., \& Kenney, J. D.~P. 2004, The Astrophysical Journal, 613, 851,
  \dodoi{10.1086/423190}

\bibitem[{{Kreckel} {et~al.}(2016){Kreckel}, {Blanc}, {Schinnerer}, {Groves},
  {Adamo}, {Hughes}, \& {Meidt}}]{Kreckel16}
{Kreckel}, K., {Blanc}, G.~A., {Schinnerer}, E., {et~al.} 2016, \apj, 827, 103,
  \dodoi{10.3847/0004-637X/827/2/103}

\bibitem[{{Kumari} {et~al.}(2019){Kumari}, {Maiolino}, {Belfiore}, \&
  {Curti}}]{Kumari19}
{Kumari}, N., {Maiolino}, R., {Belfiore}, F., \& {Curti}, M. 2019, \mnras, 485,
  367, \dodoi{10.1093/mnras/stz366}

\bibitem[{{Lacerda} {et~al.}(2018){Lacerda}, {Cid Fernandes}, {Couto},
  {Stasi{\'n}ska}, {Garc{\'\i}a-Benito}, {Vale Asari}, {P{\'e}rez},
  {Gonz{\'a}lez Delgado}, {S{\'a}nchez}, \& {de Amorim}}]{Lacerda18}
{Lacerda}, E.~A.~D., {Cid Fernandes}, R., {Couto}, G.~S., {et~al.} 2018,
  \mnras, 474, 3727, \dodoi{10.1093/mnras/stx3022}

\bibitem[{{Levy} {et~al.}(2019){Levy}, {Bolatto}, {S{\'a}nchez}, {Blitz},
  {Colombo}, {Kalinova}, {L{\'o}pez-Cob{\'a}}, {Ostriker}, {Teuben}, {Utomo},
  {Vogel}, \& {Wong}}]{Levy19}
{Levy}, R.~C., {Bolatto}, A.~D., {S{\'a}nchez}, S.~F., {et~al.} 2019, \apj,
  882, 84, \dodoi{10.3847/1538-4357/ab2ed4}

\bibitem[{{Madsen} {et~al.}(2006){Madsen}, {Reynolds}, \& {Haffner}}]{Madsen06}
{Madsen}, G.~J., {Reynolds}, R.~J., \& {Haffner}, L.~M. 2006, \apj, 652, 401,
  \dodoi{10.1086/508441}

\bibitem[{{Maier} {et~al.}(2006){Maier}, {Lilly}, {Carollo}, {Meisenheimer},
  {Hippelein}, \& {Stockton}}]{Maier06}
{Maier}, C., {Lilly}, S.~J., {Carollo}, C.~M., {et~al.} 2006, \apj, 639, 858,
  \dodoi{10.1086/499518}

\bibitem[{{Markwardt}(2009{\natexlab{a}})}]{MPFIT09}
{Markwardt}, C.~B. 2009{\natexlab{a}}, in Astronomical Society of the Pacific
  Conference Series, Vol. 411, Astronomical Data Analysis Software and Systems
  XVIII, ed. D.~A. {Bohlender}, D.~{Durand}, \& P.~{Dowler}, 251.
\newblock \doarXiv{0902.2850}

\bibitem[{{Markwardt}(2009{\natexlab{b}})}]{Markwardt09}
{Markwardt}, C.~B. 2009{\natexlab{b}}, in Astronomical Society of the Pacific
  Conference Series, Vol. 411, Astronomical Data Analysis Software and Systems
  XVIII, ed. D.~A. {Bohlender}, D.~{Durand}, \& P.~{Dowler}, 251.
\newblock \doarXiv{0902.2850}

\bibitem[{{Martin}(1997)}]{Martin91}
{Martin}, C.~L. 1997, \apj, 491, 561, \dodoi{10.1086/304978}

\bibitem[{{Mingozzi} {et~al.}(2020){Mingozzi}, {Belfiore}, {Cresci}, {Bundy},
  {Bershady}, {Bizyaev}, {Blanc}, {Boquien}, {Drory}, {Fu}, {Maiolino},
  {Riffel}, {Schaefer}, {Storchi-Bergmann}, {Telles}, {Tremonti}, {Zakamska},
  \& {Zhang}}]{Mingozzi20}
{Mingozzi}, M., {Belfiore}, F., {Cresci}, G., {et~al.} 2020, \aap, 636, A42,
  \dodoi{10.1051/0004-6361/201937203}

\bibitem[{{Minter} \& {Balser}(1998)}]{Minter98}
{Minter}, A., \& {Balser}, D.~S. 1998, {Turbulent Heating in the Galactic
  Diffuse Ionized Gas}, ed. D.~{Breitschwerdt}, M.~J. {Freyberg}, \&
  J.~{Truemper}, Vol. 506, 543--546, \dodoi{10.1007/BFb0104779}

\bibitem[{{Minter} \& {Spangler}(1997)}]{Minter97}
{Minter}, A.~H., \& {Spangler}, S.~R. 1997, \apj, 485, 182,
  \dodoi{10.1086/304396}

\bibitem[{{M{\"u}ller} {et~al.}(2021){M{\"u}ller}, {Poggianti}, {Pfrommer},
  {Adebahr}, {Serra}, {Ignesti}, {Sparre}, {Gitti}, {Dettmar}, {Vulcani}, \&
  {Moretti}}]{Muller21}
{M{\"u}ller}, A., {Poggianti}, B.~M., {Pfrommer}, C., {et~al.} 2021, Nature
  Astronomy, 5, 159, \dodoi{10.1038/s41550-020-01234-7}

\bibitem[{{Nagao} {et~al.}(2006){Nagao}, {Maiolino}, \& {Marconi}}]{Nagao06}
{Nagao}, T., {Maiolino}, R., \& {Marconi}, A. 2006, \aap, 459, 85,
  \dodoi{10.1051/0004-6361:20065216}

\bibitem[{{Oey} {et~al.}(2007){Oey}, {Meurer}, {Yelda}, {Furst},
  {Caballero-Nieves}, {Hanish}, {Levesque}, {Thilker}, {Walth},
  {Bland-Hawthorn}, {Dopita}, {Ferguson}, {Heckman}, {Doyle}, {Drinkwater},
  {Freeman}, {Kennicutt}, {Kilborn}, {Knezek}, {Koribalski}, {Meyer}, {Putman},
  {Ryan-Weber}, {Smith}, {Staveley-Smith}, {Webster}, {Werk}, \&
  {Zwaan}}]{Oey07}
{Oey}, M.~S., {Meurer}, G.~R., {Yelda}, S., {et~al.} 2007, \apj, 661, 801,
  \dodoi{10.1086/517867}

\bibitem[{{Osterbrock} \& {Martel}(1992)}]{Osterbrock92}
{Osterbrock}, D.~E., \& {Martel}, A. 1992, \pasp, 104, 76,
  \dodoi{10.1086/132961}

\bibitem[{{Otte} {et~al.}(2002){Otte}, {Gallagher}, \& {Reynolds}}]{Otte02}
{Otte}, B., {Gallagher}, J.~S., I., \& {Reynolds}, R.~J. 2002, \apj, 572, 823,
  \dodoi{10.1086/340381}

\bibitem[{{Poetrodjojo} {et~al.}(2019){Poetrodjojo}, {D'Agostino}, {Groves},
  {Kewley}, {Ho}, {Rich}, {Madore}, \& {Seibert}}]{Poetrodjojo19}
{Poetrodjojo}, H., {D'Agostino}, J.~J., {Groves}, B., {et~al.} 2019, \mnras,
  487, 79, \dodoi{10.1093/mnras/stz1241}

\bibitem[{{Poetrodjojo} {et~al.}(2018){Poetrodjojo}, {Groves}, {Kewley},
  {Medling}, {Sweet}, {van de Sande}, {Sanchez}, {Bland-Hawthorn}, {Brough},
  {Bryant}, {Cortese}, {Croom}, {L{\'o}pez-S{\'a}nchez}, {Richards}, {Zafar},
  {Lawrence}, {Lorente}, {Owers}, \& {Scott}}]{Poetrodjojo18}
{Poetrodjojo}, H., {Groves}, B., {Kewley}, L.~J., {et~al.} 2018, \mnras, 479,
  5235, \dodoi{10.1093/mnras/sty1782}

\bibitem[{{Poggianti} {et~al.}(2017){Poggianti}, {Moretti}, {Gullieuszik},
  {Fritz}, {Jaff{\'e}}, {Bettoni}, {Fasano}, {Bellhouse}, {Hau}, {Vulcani},
  {Biviano}, {Omizzolo}, {Paccagnella}, {D'Onofrio}, {Cava}, {Sheen}, {Couch},
  \& {Owers}}]{Poggianti17}
{Poggianti}, B.~M., {Moretti}, A., {Gullieuszik}, M., {et~al.} 2017, \apj, 844,
  48, \dodoi{10.3847/1538-4357/aa78ed}

\bibitem[{{Poggianti} {et~al.}(2019{\natexlab{a}}){Poggianti}, {Ignesti},
  {Gitti}, {Wolter}, {Brighenti}, {Biviano}, {George}, {Vulcani},
  {Gullieuszik}, {Moretti}, {Paladino}, {Bettoni}, {Franchetto}, {Jaff{\'e}},
  {Radovich}, {Roediger}, {Tomi{\v{c}}i{\'c}}, {Tonnesen}, {Bellhouse},
  {Fritz}, \& {Omizzolo}}]{Poggianti19b}
{Poggianti}, B.~M., {Ignesti}, A., {Gitti}, M., {et~al.} 2019{\natexlab{a}},
  \apj, 887, 155, \dodoi{10.3847/1538-4357/ab5224}

\bibitem[{{Poggianti} {et~al.}(2019{\natexlab{b}}){Poggianti}, {Gullieuszik},
  {Tonnesen}, {Moretti}, {Vulcani}, {Radovich}, {Jaff{\'e}}, {Fritz},
  {Bettoni}, {Franchetto}, {Fasano}, {Bellhouse}, \& {Omizzolo}}]{Poggianti19}
{Poggianti}, B.~M., {Gullieuszik}, M., {Tonnesen}, S., {et~al.}
  2019{\natexlab{b}}, \mnras, 482, 4466, \dodoi{10.1093/mnras/sty2999}

\bibitem[{{Raymond}(1992)}]{Raymond92}
{Raymond}, J.~C. 1992, \apj, 384, 502, \dodoi{10.1086/170892}

\bibitem[{{Rela{\~n}o} {et~al.}(2012){Rela{\~n}o}, {Kennicutt}, {Eldridge},
  {Lee}, \& {Verley}}]{Relano12}
{Rela{\~n}o}, M., {Kennicutt}, R.~C., J., {Eldridge}, J.~J., {Lee}, J.~C., \&
  {Verley}, S. 2012, \mnras, 423, 2933,
  \dodoi{10.1111/j.1365-2966.2012.21107.x}

\bibitem[{{Reynolds}(1984)}]{Reynolds84}
{Reynolds}, R.~J. 1984, \apj, 282, 191, \dodoi{10.1086/162190}

\bibitem[{{Reynolds} \& {Cox}(1992)}]{Reynolds92}
{Reynolds}, R.~J., \& {Cox}, D.~P. 1992, \apjl, 400, L33,
  \dodoi{10.1086/186642}

\bibitem[{{Reynolds} {et~al.}(2001){Reynolds}, {Sterling}, {Haffner}, \&
  {Tufte}}]{Reynolds01}
{Reynolds}, R.~J., {Sterling}, N.~C., {Haffner}, L.~M., \& {Tufte}, S.~L. 2001,
  \apjl, 548, L221, \dodoi{10.1086/319119}

\bibitem[{{S{\'a}nchez}(2020)}]{Sanchez20}
{S{\'a}nchez}, S.~F. 2020, \araa, 58, 99,
  \dodoi{10.1146/annurev-astro-012120-013326}

\bibitem[{{S{\'a}nchez} {et~al.}(2012){S{\'a}nchez}, {Rosales-Ortega},
  {Marino}, {Iglesias-P{\'a}ramo}, {V{\'\i}lchez}, {Kennicutt}, {D{\'\i}az},
  {Mast}, {Monreal-Ibero}, {Garc{\'\i}a-Benito}, {Bland-Hawthorn}, {P{\'e}rez},
  {Gonz{\'a}lez Delgado}, {Husemann}, {L{\'o}pez-S{\'a}nchez}, {Cid Fernandes},
  {Kehrig}, {Walcher}, {Gil de Paz}, \& {Ellis}}]{Sanchez12}
{S{\'a}nchez}, S.~F., {Rosales-Ortega}, F.~F., {Marino}, R.~A., {et~al.} 2012,
  \aap, 546, A2, \dodoi{10.1051/0004-6361/201219578}

\bibitem[{{S{\'a}nchez} {et~al.}(2014){S{\'a}nchez}, {Rosales-Ortega},
  {Iglesias-P{\'a}ramo}, {Moll{\'a}}, {Barrera-Ballesteros}, {Marino},
  {P{\'e}rez}, {S{\'a}nchez-Blazquez}, {Gonz{\'a}lez Delgado}, {Cid Fernandes},
  {de Lorenzo-C{\'a}ceres}, {Mendez-Abreu}, {Galbany}, {Falcon-Barroso},
  {Miralles-Caballero}, {Husemann}, {Garc{\'\i}a-Benito}, {Mast}, {Walcher},
  {Gil de Paz}, {Garc{\'\i}a-Lorenzo}, {Jungwiert}, {V{\'\i}lchez},
  {J{\'\i}lkov{\'a}}, {Lyubenova}, {Cortijo-Ferrero}, {D{\'\i}az}, {Wisotzki},
  {M{\'a}rquez}, {Bland-Hawthorn}, {Ellis}, {van de Ven}, {Jahnke},
  {Papaderos}, {Gomes}, {Mendoza}, \& {L{\'o}pez-S{\'a}nchez}}]{Sanchez14}
{S{\'a}nchez}, S.~F., {Rosales-Ortega}, F.~F., {Iglesias-P{\'a}ramo}, J.,
  {et~al.} 2014, \aap, 563, A49, \dodoi{10.1051/0004-6361/201322343}

\bibitem[{{S{\'a}nchez} {et~al.}(2015){S{\'a}nchez}, {Galbany}, {P{\'e}rez},
  {S{\'a}nchez-Bl{\'a}zquez}, {Falc{\'o}n-Barroso}, {Rosales-Ortega},
  {S{\'a}nchez-Menguiano}, {Marino}, {Kuncarayakti}, {Anderson}, {Kruehler},
  {Cano-D{\'\i}az}, {Barrera-Ballesteros}, \&
  {Gonz{\'a}lez-Gonz{\'a}lez}}]{Sanchez15}
{S{\'a}nchez}, S.~F., {Galbany}, L., {P{\'e}rez}, E., {et~al.} 2015, \aap, 573,
  A105, \dodoi{10.1051/0004-6361/201424950}

\bibitem[{{S{\'a}nchez-Menguiano} {et~al.}(2018){S{\'a}nchez-Menguiano},
  {S{\'a}nchez}, {P{\'e}rez}, {Ruiz-Lara}, {Galbany}, {Anderson},
  {Kr{\"u}hler}, {Kuncarayakti}, \& {Lyman}}]{SanchezMeng2018}
{S{\'a}nchez-Menguiano}, L., {S{\'a}nchez}, S.~F., {P{\'e}rez}, I., {et~al.}
  2018, \aap, 609, A119, \dodoi{10.1051/0004-6361/201731486}

\bibitem[{{Sanders} {et~al.}(2017){Sanders}, {Shapley}, {Zhang}, \&
  {Yan}}]{Sanders17}
{Sanders}, R.~L., {Shapley}, A.~E., {Zhang}, K., \& {Yan}, R. 2017, \apj, 850,
  136, \dodoi{10.3847/1538-4357/aa93e4}

\bibitem[{{Scaife}(2013)}]{Scaife13}
{Scaife}, A. M.~M. 2013, Advances in Astronomy, 2013, 390287,
  \dodoi{10.1155/2013/390287}

\bibitem[{{Schlafly} \& {Finkbeiner}(2011)}]{Schlafly11}
{Schlafly}, E.~F., \& {Finkbeiner}, D.~P. 2011, \apj, 737, 103,
  \dodoi{10.1088/0004-637X/737/2/103}

\bibitem[{{Schlegel} {et~al.}(1998){Schlegel}, {Finkbeiner}, \& {Davis}}]{SFD}
{Schlegel}, D.~J., {Finkbeiner}, D.~P., \& {Davis}, M. 1998, \apj, 500, 525,
  \dodoi{10.1086/305772}

\bibitem[{{Searle}(1971)}]{Searle71}
{Searle}, L. 1971, \apj, 168, 327, \dodoi{10.1086/151090}

\bibitem[{{Simpson} {et~al.}(2007){Simpson}, {Colgan}, {Cotera}, {Erickson},
  {Hollenbach}, {Kaufman}, \& {Rubin}}]{Simpson07}
{Simpson}, J.~P., {Colgan}, S. W.~J., {Cotera}, A.~S., {et~al.} 2007, \apj,
  670, 1115, \dodoi{10.1086/522295}

\bibitem[{{Singh} {et~al.}(2013){Singh}, {van de Ven}, {Jahnke}, {Lyubenova},
  {Falc{\'o}n-Barroso}, {Alves}, {Cid Fernandes}, {Galbany},
  {Garc{\'\i}a-Benito}, {Husemann}, {Kennicutt}, {Marino}, {M{\'a}rquez},
  {Masegosa}, {Mast}, {Pasquali}, {S{\'a}nchez}, {Walcher}, {Wild}, {Wisotzki},
  \& {Ziegler}}]{Singh13}
{Singh}, R., {van de Ven}, G., {Jahnke}, K., {et~al.} 2013, \aap, 558, A43,
  \dodoi{10.1051/0004-6361/201322062}

\bibitem[{{Slavin} {et~al.}(1993){Slavin}, {Shull}, \& {Begelman}}]{Slavin93}
{Slavin}, J.~D., {Shull}, J.~M., \& {Begelman}, M.~C. 1993, \apj, 407, 83,
  \dodoi{10.1086/172494}

\bibitem[{{Sparre} {et~al.}(2020){Sparre}, {Pfrommer}, \& {Ehlert}}]{Sparre20}
{Sparre}, M., {Pfrommer}, C., \& {Ehlert}, K. 2020, \mnras, 499, 4261,
  \dodoi{10.1093/mnras/staa3177}

\bibitem[{{Tomi{\v{c}}i{\'c}} {et~al.}(2017){Tomi{\v{c}}i{\'c}}, {Kreckel},
  {Groves}, {Schinnerer}, {Sandstrom}, {Kapala}, {Blanc}, \&
  {Leroy}}]{Tomicic17}
{Tomi{\v{c}}i{\'c}}, N., {Kreckel}, K., {Groves}, B., {et~al.} 2017, \apj, 844,
  155, \dodoi{10.3847/1538-4357/aa7b30}

\bibitem[{{Tomi{\v{c}}i{\'c}} {et~al.}(2021){Tomi{\v{c}}i{\'c}}, {Vulcani},
  {Poggianti}, {Mingozzi}, {Werle}, {Bettoni}, {Franchetto}, {Gullieuszik},
  {Moretti}, {Fritz}, \& {Bellhouse}}]{Tomicic21a}
{Tomi{\v{c}}i{\'c}}, N., {Vulcani}, B., {Poggianti}, B.~M., {et~al.} 2021,
  \apj, 907, 22, \dodoi{10.3847/1538-4357/abca93}

\bibitem[{{Toomre} \& {Toomre}(1972)}]{Toomre72}
{Toomre}, A., \& {Toomre}, J. 1972, \apj, 178, 623, \dodoi{10.1086/151823}

\bibitem[{{Tremonti} {et~al.}(2004){Tremonti}, {Heckman}, {Kauffmann},
  {Brinchmann}, {Charlot}, {White}, {Seibert}, {Peng}, {Schlegel}, {Uomoto},
  {Fukugita}, \& {Brinkmann}}]{Tremonti04}
{Tremonti}, C.~A., {Heckman}, T.~M., {Kauffmann}, G., {et~al.} 2004, \apj, 613,
  898, \dodoi{10.1086/423264}

\bibitem[{{Vale Asari} {et~al.}(2019){Vale Asari}, {Couto}, {Cid Fernandes},
  {Stasi{\'n}ska}, {de Amorim}, {Ruschel-Dutra}, {Werle}, \&
  {Florido}}]{Asari19}
{Vale Asari}, N., {Couto}, G.~S., {Cid Fernandes}, R., {et~al.} 2019, \mnras,
  489, 4721, \dodoi{10.1093/mnras/stz2470}

\bibitem[{{Vila-Costas} \& {Edmunds}(1992)}]{Costas92}
{Vila-Costas}, M.~B., \& {Edmunds}, M.~G. 1992, \mnras, 259, 121,
  \dodoi{10.1093/mnras/259.1.121}

\bibitem[{{Vogt} {et~al.}(2015){Vogt}, {Dopita}, {Borthakur},
  {Verdes-Montenegro}, {Heckman}, {Yun}, \& {Chambers}}]{Vogt15}
{Vogt}, F. P.~A., {Dopita}, M.~A., {Borthakur}, S., {et~al.} 2015, \mnras, 450,
  2593, \dodoi{10.1093/mnras/stv749}

\bibitem[{{Vulcani} {et~al.}(2017){Vulcani}, {Moretti}, {Poggianti}, {Fasano},
  {Fritz}, {Gullieuszik}, {Duc}, {Jaff{\'e}}, \& {Bettoni}}]{Vulcani17}
{Vulcani}, B., {Moretti}, A., {Poggianti}, B.~M., {et~al.} 2017, \apj, 850,
  163, \dodoi{10.3847/1538-4357/aa9652}

\bibitem[{{Vulcani} {et~al.}(2018{\natexlab{a}}){Vulcani}, {Poggianti},
  {Gullieuszik}, {Moretti}, {Tonnesen}, {Jaff{\'e}}, {Fritz}, {Fasano}, \&
  {Bettoni}}]{Vulcani18b}
{Vulcani}, B., {Poggianti}, B.~M., {Gullieuszik}, M., {et~al.}
  2018{\natexlab{a}}, \apjl, 866, L25, \dodoi{10.3847/2041-8213/aae68b}

\bibitem[{{Vulcani} {et~al.}(2018{\natexlab{b}}){Vulcani}, {Poggianti},
  {Moretti}, {Mapelli}, {Fasano}, {Fritz}, {Jaff{\'e}}, {Bettoni},
  {Gullieuszik}, \& {Bellhouse}}]{Vulcani18}
{Vulcani}, B., {Poggianti}, B.~M., {Moretti}, A., {et~al.} 2018{\natexlab{b}},
  \apj, 852, 94, \dodoi{10.3847/1538-4357/aa992c}

\bibitem[{{Vulcani} {et~al.}(2018{\natexlab{c}}){Vulcani}, {Poggianti},
  {Jaff{\'e}}, {Moretti}, {Fritz}, {Gullieuszik}, {Bettoni}, {Fasano},
  {Tonnesen}, \& {McGee}}]{Vulcani18a}
{Vulcani}, B., {Poggianti}, B.~M., {Jaff{\'e}}, Y.~L., {et~al.}
  2018{\natexlab{c}}, \mnras, 480, 3152, \dodoi{10.1093/mnras/sty2095}

\bibitem[{{Vulcani} {et~al.}(2019{\natexlab{a}}){Vulcani}, {Poggianti},
  {Moretti}, {Gullieuszik}, {Fritz}, {Franchetto}, {Fasano}, {Bettoni}, \&
  {Jaff{\'e}}}]{Vulcani19a}
{Vulcani}, B., {Poggianti}, B.~M., {Moretti}, A., {et~al.} 2019{\natexlab{a}},
  \mnras, 487, 2278, \dodoi{10.1093/mnras/stz1399}

\bibitem[{{Vulcani} {et~al.}(2019{\natexlab{b}}){Vulcani}, {Poggianti},
  {Moretti}, {Franchetto}, {Gullieuszik}, {Fritz}, {Bettoni}, {Tonnesen},
  {Radovich}, {Jaff{\'e}}, {McGee}, {Bellhouse}, \& {Fasano}}]{Vulcani19}
---. 2019{\natexlab{b}}, \mnras, 488, 1597, \dodoi{10.1093/mnras/stz1829}

\bibitem[{{Vulcani} {et~al.}(2020){Vulcani}, {Poggianti}, {Tonnesen}, {McGee},
  {Moretti}, {Fritz}, {Gullieuszik}, {Jaffe}, {Franchetto}, {Tomicic},
  {Mingozzi}, {Bettoni}, \& {Wolter}}]{Vulcani20b}
{Vulcani}, B., {Poggianti}, B.~M., {Tonnesen}, S., {et~al.} 2020, arXiv
  e-prints, arXiv:2007.04996.
\newblock \doarXiv{2007.04996}

\bibitem[{{Vulcani} {et~al.}(2021){Vulcani}, {Poggianti}, {Moretti},
  {Franchetto}, {Bacchini}, {McGee}, {Jaffe}, {Mingozzi}, {Werle}, {Tomicic},
  {Fritz}, {Bettoni}, {Wolter}, \& {Gullieuszik}}]{Vulcani2021}
{Vulcani}, B., {Poggianti}, B.~M., {Moretti}, A., {et~al.} 2021, arXiv
  e-prints, arXiv:2104.02089.
\newblock \doarXiv{2104.02089}

\bibitem[{{Weingartner} \& {Draine}(2001)}]{Weingartner01}
{Weingartner}, J.~C., \& {Draine}, B.~T. 2001, \apjs, 134, 263,
  \dodoi{10.1086/320852}

\bibitem[{{Werle} {et~al.}(2020){Werle}, {Cid Fernandes}, {Vale Asari},
  {Coelho}, {Bruzual}, {Charlot}, {de Carvalho}, {Herpich}, {Mendes de
  Oliveira}, {Sodr{\'e}}, {Ruschel-Dutra}, {de Amorim}, \& {Sampaio}}]{Werle20}
{Werle}, A., {Cid Fernandes}, R., {Vale Asari}, N., {et~al.} 2020, \mnras, 497,
  3251, \dodoi{10.1093/mnras/staa2217}

\bibitem[{Yeh \& Matzner(2012)}]{Yeh12}
Yeh, S. C.~C., \& Matzner, C.~D. 2012, The Astrophysical Journal, 757, 108,
  \dodoi{10.1088/0004-637x/757/2/108}

\bibitem[{{Zhang} {et~al.}(2017){Zhang}, {Yan}, {Bundy}, {Bershady}, {Haffner},
  {Walterbos}, {Maiolino}, {Tremonti}, {Thomas}, {Drory}, {Jones}, {Belfiore},
  {S{\'a}nchez}, {Diamond-Stanic}, {Bizyaev}, {Nitschelm}, {Andrews},
  {Brinkmann}, {Brownstein}, {Cheung}, {Li}, {Law}, {Roman Lopes}, {Oravetz},
  {Pan}, {Storchi Bergmann}, \& {Simmons}}]{Zhang17}
{Zhang}, K., {Yan}, R., {Bundy}, K., {et~al.} 2017, \mnras, 466, 3217,
  \dodoi{10.1093/mnras/stw3308}

\end{thebibliography}

\end{document}